\documentclass{webofc}
\usepackage{subcaption}
\usepackage{style}
\usepackage[varg]{txfonts}   
\usepackage{hyperref}
\usepackage{url}
\hypersetup{colorlinks=true,citecolor=blue,urlcolor=blue,linkcolor=blue}
\begin{document}
\title{Optimizing Cell-Based Negative Weight Mitigation\\ with Optimal Transport} 

\author{\firstname{Lauren} \lastname{Hay}\inst{1}\fnsep\thanks{\email{lauren_hay@brown.edu}} \and
        \firstname{Rishabh} \lastname{Jain}\inst{1}\fnsep \and
        \firstname{Matt} \lastname{LeBlanc}\inst{1}\fnsep \and
        \firstname{Jennifer} \lastname{Roloff}\inst{1}\fnsep}

\institute{Brown University
          }
\abstract{As the accuracy of experimental results in high energy physics (HEP) increases, so does the demand for precision Monte Carlo (MC) simulation. Higher-accuracy event generation at next-to-leading order (NLO) and beyond brings with it negatively weighted events. These negatively weighted events reduce the statistical power of samples, increasing the number of events that need to be produced and straining already limited computational resources. We present a post-hoc reweighting scheme that employs cell-based resampling using an IRC-safe metric to define the cell radii. An optimal metric embeds kinematically similar events within the same cells, minimizing bias when reweighted. This motivates our exploration of a Optimal Transport (OT) based distance metrics. We compare the performance of the reweighting algorithm with different choices of metric, and explicitly demonstrate the performance on simulated \zjets events produced at NLO accuracy.}
\maketitle
\section{Introduction}
\label{sec:intro}
Analysis of experimental high energy physics data is dependent on theoretical predictions, often implemented as Monte Carlo event generators. As experimental uncertainties decrease through increased understanding of detectors and greater statistics, simulation-dependent tasks such as background estimation or calibrations become a limiting factor on the precision of experimental results.

Theoretical predictions are keeping pace with experimental precision by achieving higher and higher orders of accuracy in their calculations; the techniques necessary to perform calculations at next-to-leading order (NLO) or above introduce negative weights. Negative weights cause statistical dilution; \emph{i.e.} to achieve the same cross-section in an unweighted sample as in a positive-weight-only or LO sample, one would have to produce 
\begin{equation*}
    \frac{N(\epsilon)}{N(0)}=\frac{1}{(1-2\epsilon)^2} 
    \label{eq:dilution}
\end{equation*}
times as many events to achieve the same statistical uncertainty, where $\epsilon$ is the fraction of negative events.

The presence of negative weights can have a large impact in the context of an  experimental collaboration, with CMS citing that for a benchmark MADGRAPH AMC@NLO $t\bar{t}$ production with 23\% negative weights, removing these negative weights could reduce the needed number of MADGRAPH events by 3.4 times \cite{CMS_CDR}. ATLAS notes that vector-boson plus jets samples can reach negative weight fractions >30\%, or $>7$ times the statistical dilution~\cite{ATLAS_VplusJets}. 

As we look towards the High-Luminosity Large Hadron Collider (HL-LHC) era, experiments plan to devote more resources to these higher-precision simulation datasets, e.g. CMS expects to increase its fraction of NLO-or-higher samples from $\sim60\%$ to $\sim80\%$. Both ATLAS and CMS foresee devoting $>50\%$ of their CPU time and large portions of tape and disk to simulation, while their Phase-2 computing roadmaps indicate an imminent shortfall of computational and storage resources \cite{ATLAS_CDR, CMS_CDR}. Since experiments are expected to produce more of these computationally expensive samples, reducing negative weights becomes particularly valuable.

One promising solution to mitigate the effects of negative or pathological weights is to locally redistribute event weights via a cell-resampling algorithm applied as an afterburner to the event generation, as proposed by Andersen \emph{et al.} in Refs. \cite{Andersen:2021mvw, Andersen:2024mqh}. We propose to improve upon this method by replacing the originally-proposed distance metric with one motivated by Optimal Transport (OT), which is natively infrared and collinear (IRC) safe.

\section{Cell-resampling}
\begin{figure}[htbp]
\centering{
\includegraphics[width=0.32\textwidth]{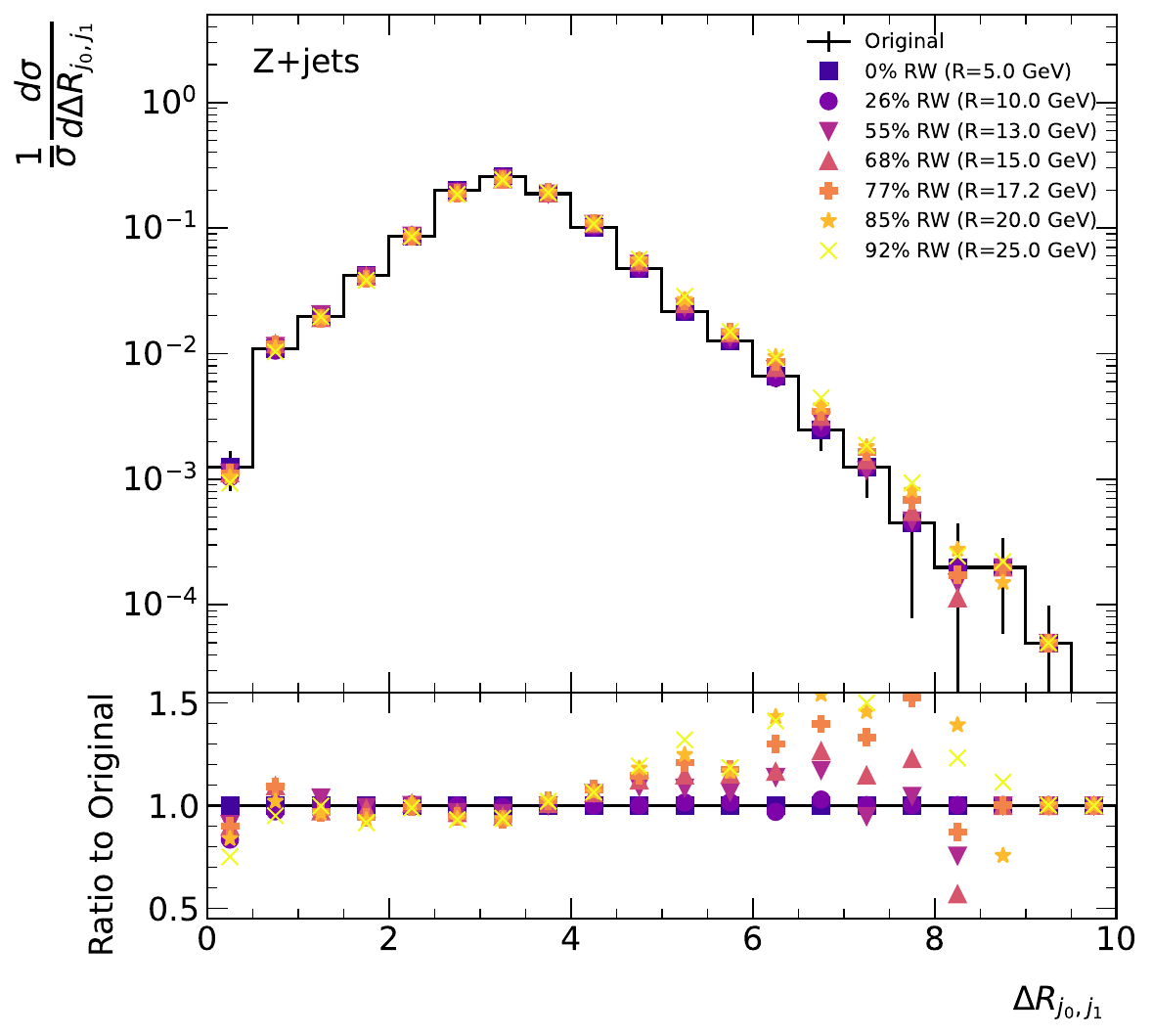}
\includegraphics[width=0.32\textwidth]{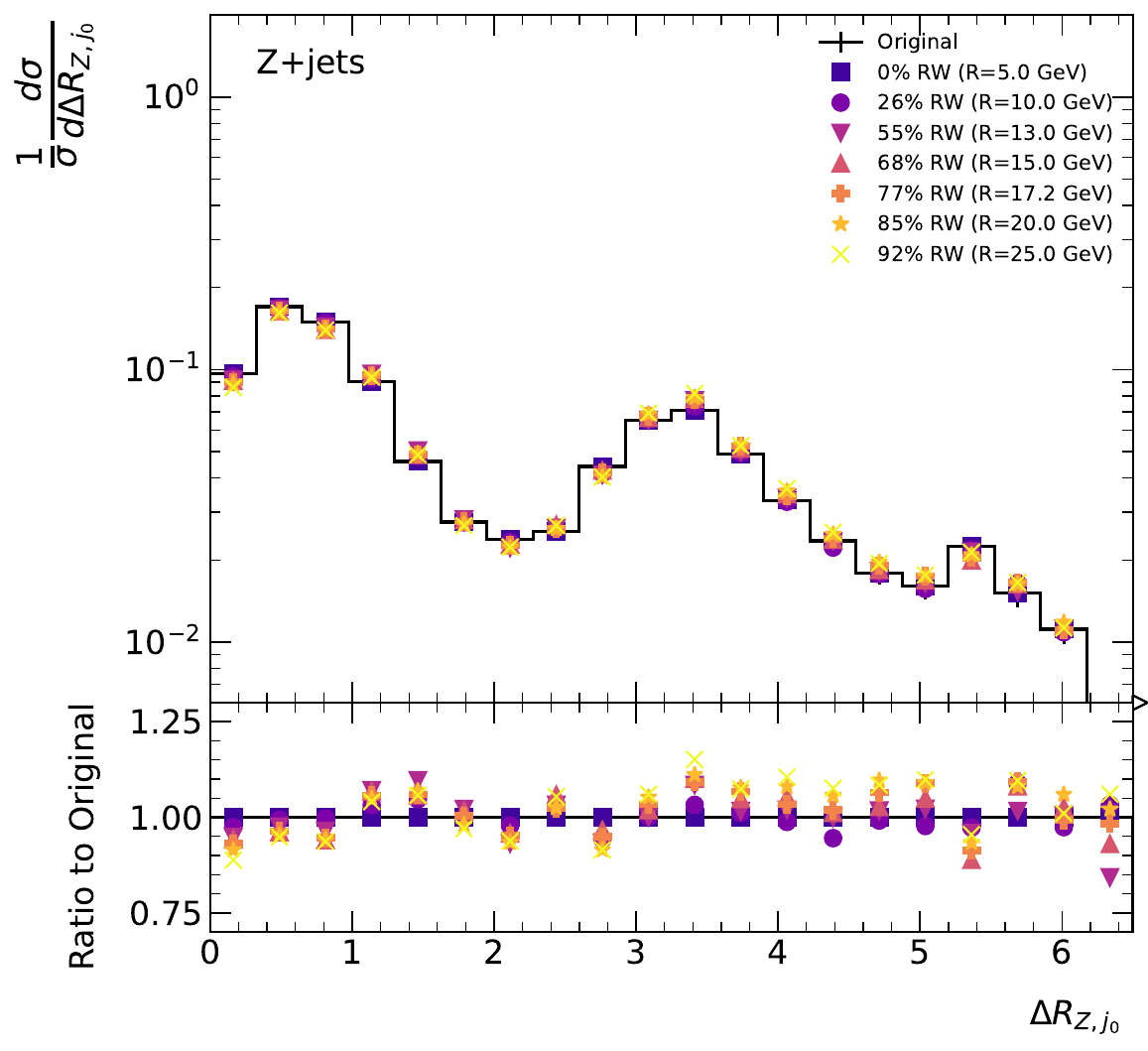}
\includegraphics[width=0.32\textwidth]{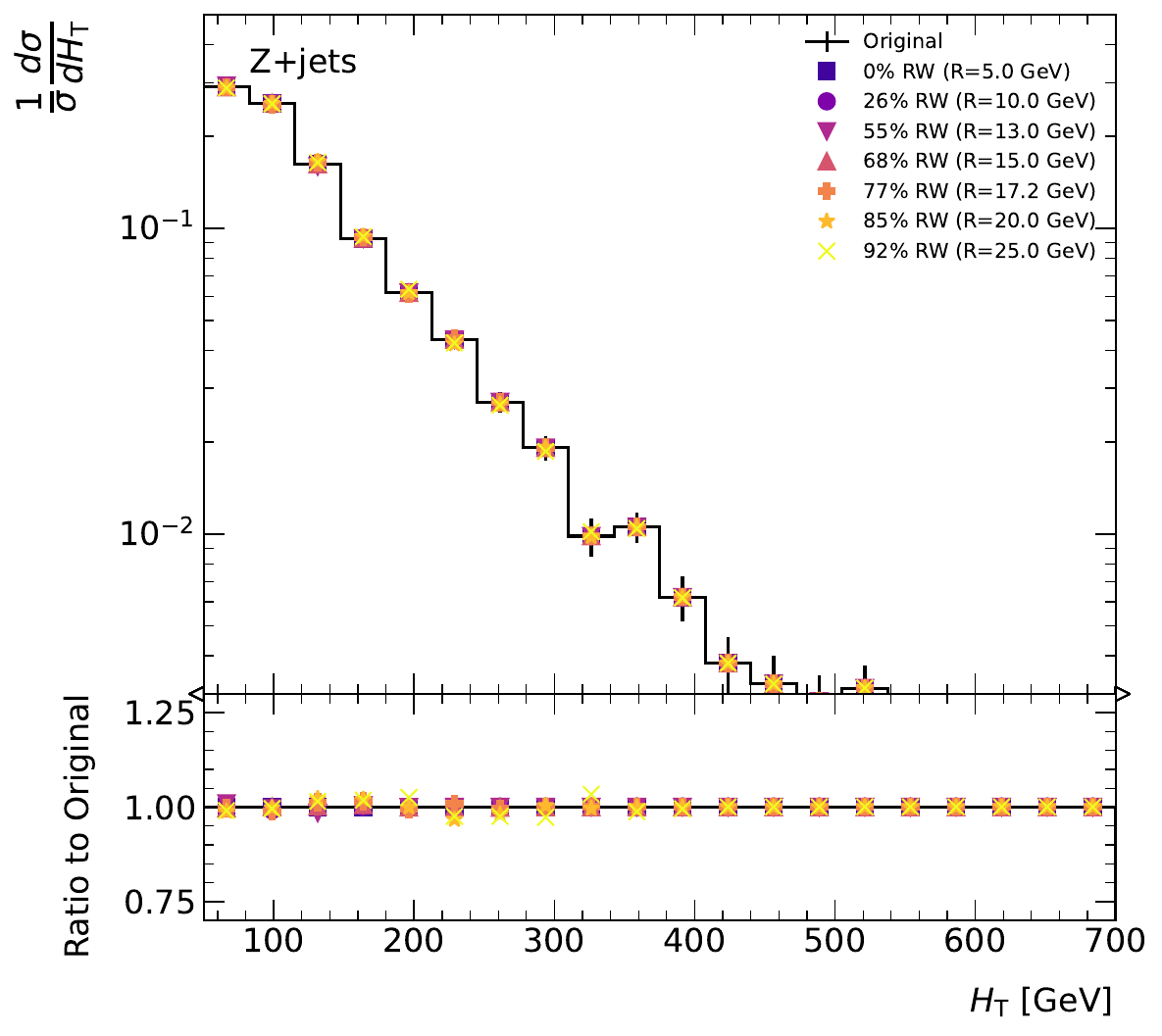} \\
\includegraphics[width=0.32\textwidth]{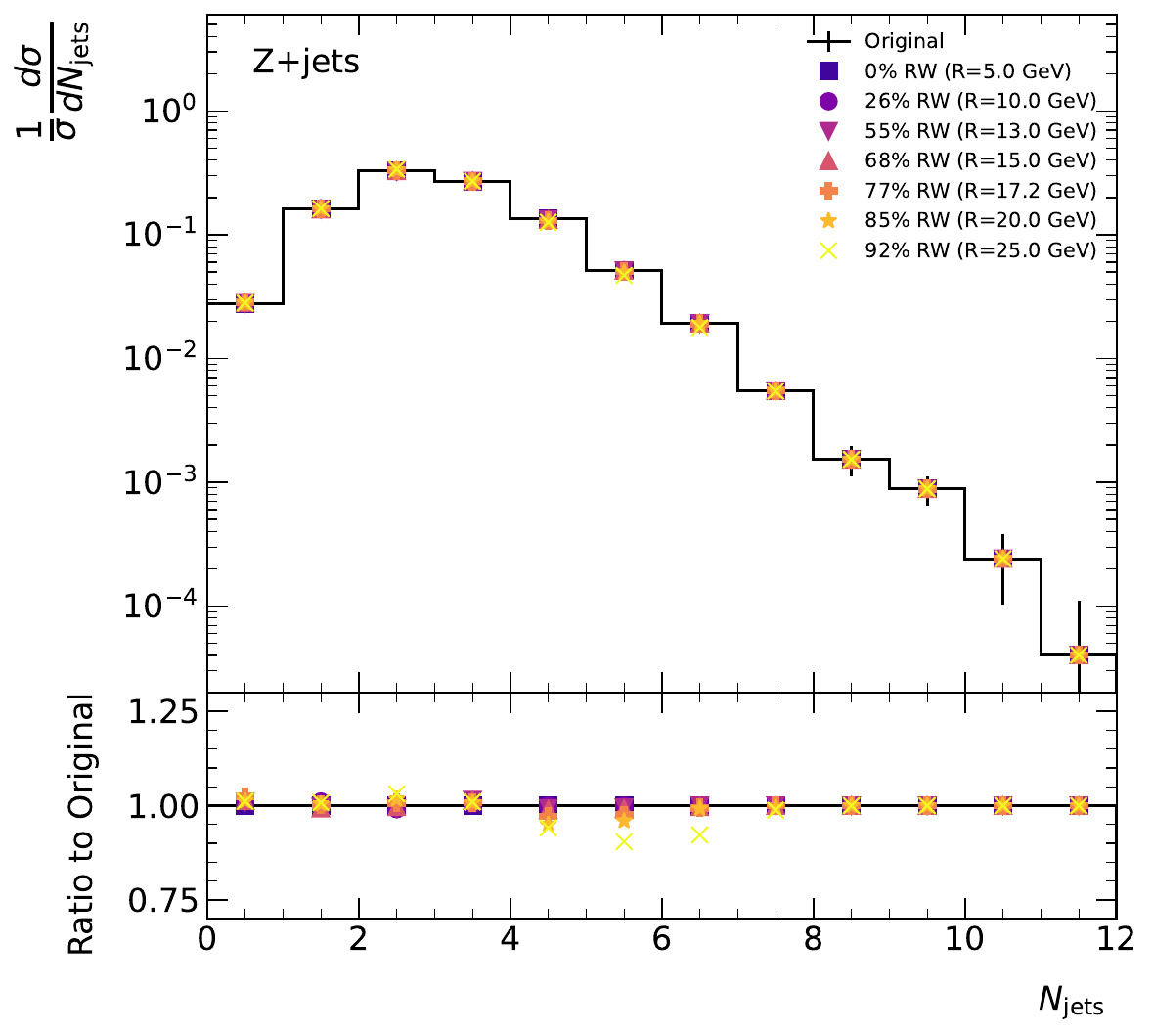}
\includegraphics[width=0.32\textwidth]{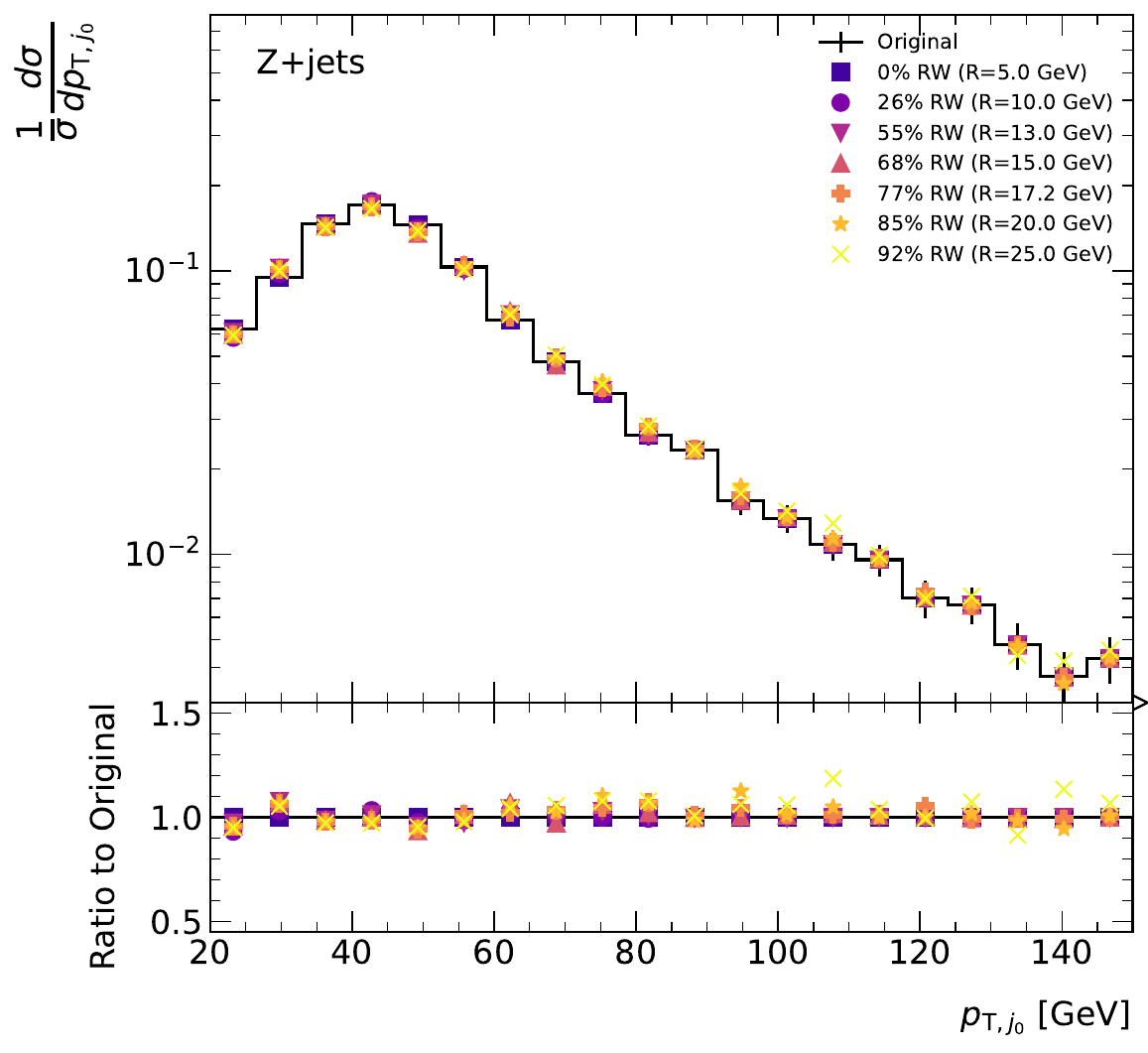}
\includegraphics[width=0.32\textwidth]{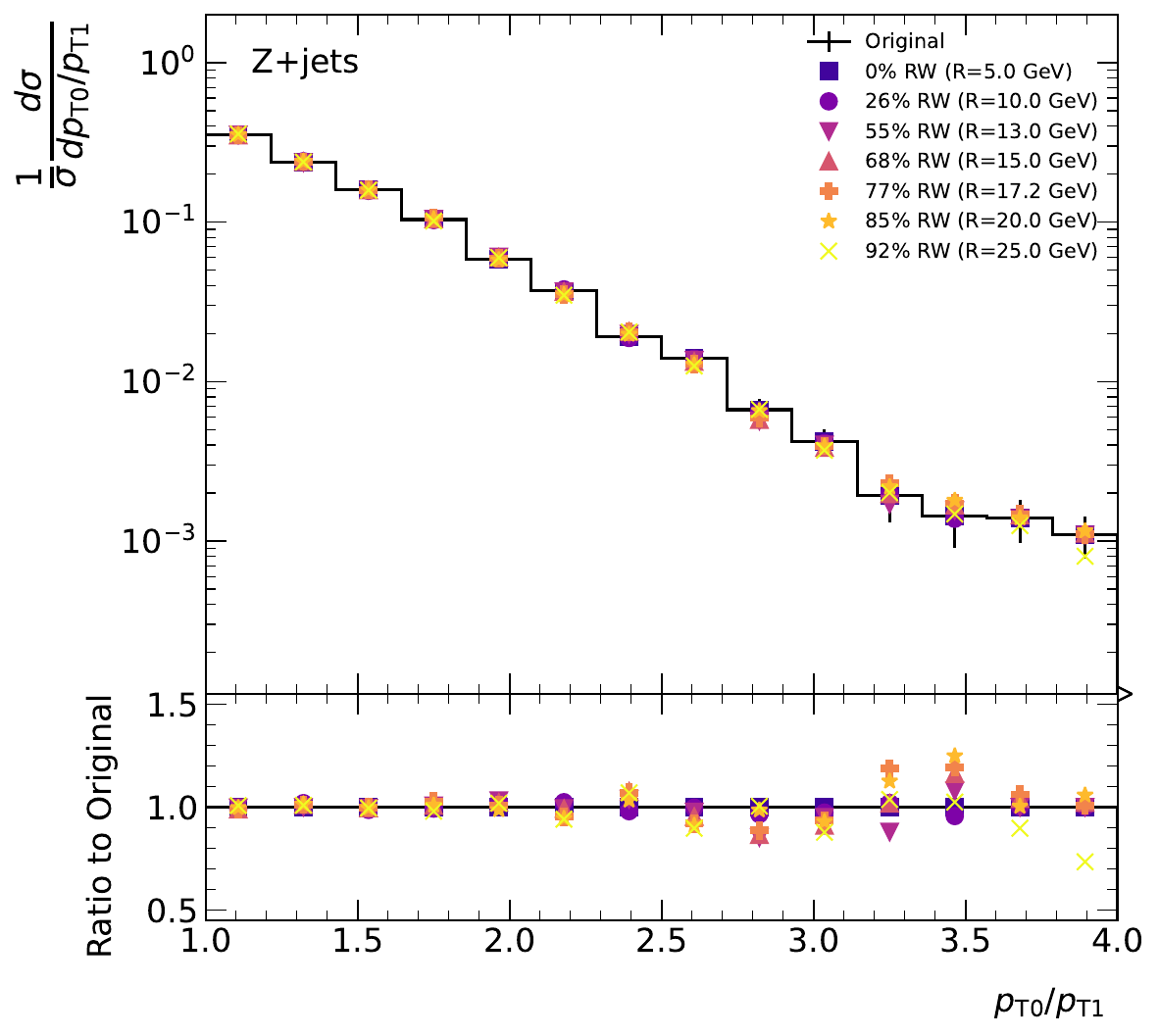}
   \caption{Comparison of the normalized hadron-level unweighted \zjets observables to samples with varying reweight (RW) fractions/maximum cell radius $R$ in the EMD metric.}
 \label{fig:fracs}
 }
\end{figure}

Cell-resampling seeks to modify event weights while preserving the total cross-section and minimizing bias in measurable observables. This is done by redistributing the weight of negatively-weighted events to positive weights nearby in phase space. A negatively weighted event is selected as the seed of a cell $C$. The cell is defined as a hypersphere of radius $R$, centered on the seed in the chosen metric space. The radius of the cell grows until the net weight of the enclosed events becomes positive. The weights inside the cell are then redefined via:
\begin{equation*}
    w_i \rightarrow \frac{\sum_{j\in C}w_j}{\sum_{j\in C}|w_j|}{|w_i|}.
\end{equation*}
\begin{figure}[!htbp]
\centering
\includegraphics[width=9cm,clip]{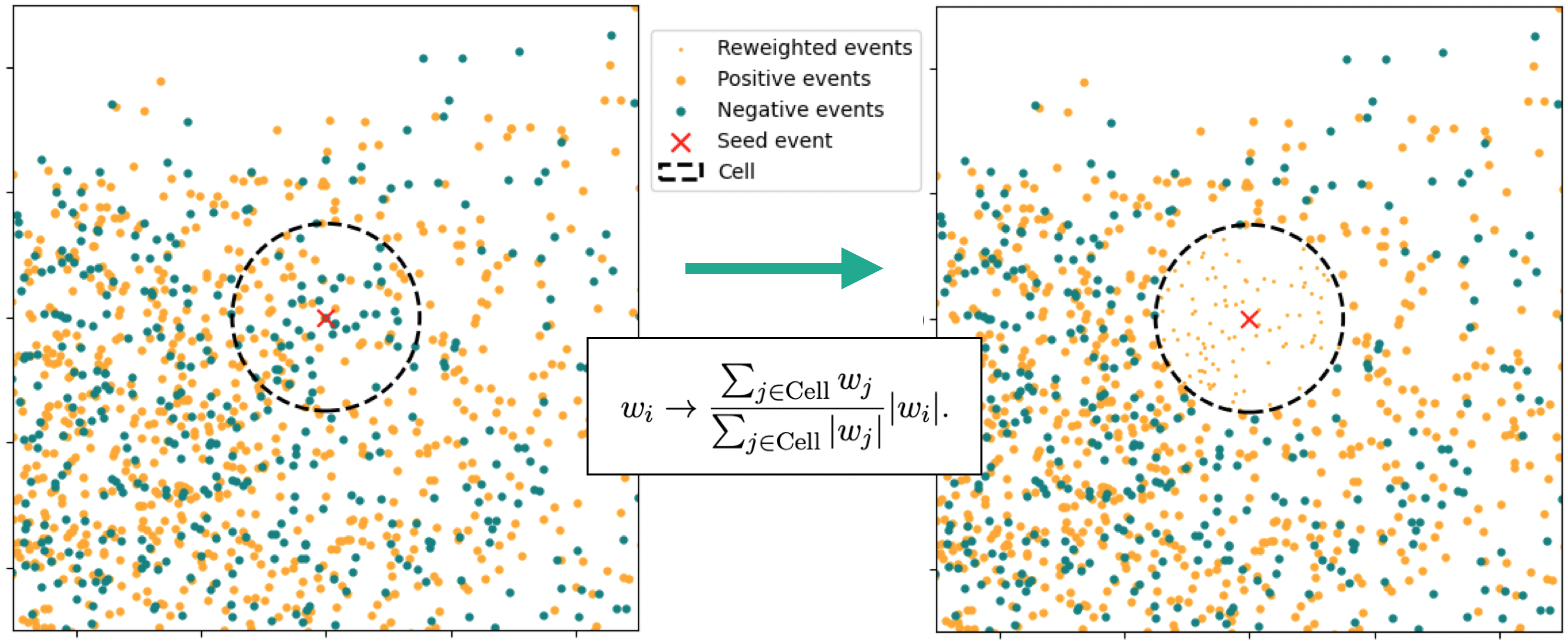}
\caption{Cartoon representation of cell-resampling in our $Z+$jets sample.}
\label{fig:cres_cartoon}       
\end{figure}
For sufficiently large samples, the phase space density becomes high enough that reweighting can be performed with a vanishingly small $R$. Kinematic bias occurs when the cross-section varies across cells, so smaller cells should result in less bias. In this proof-of-concept study, we evaluate the bias on a relatively low-statistics sample, $\mathcal{O}(10^5)$ events, roughly representative of the number of events in a single MC production job within a production campaign by an experimental collaboration.  In this case, it may be necessary to impose an upper limit on the cell size to ensure smearing effects are negligible. Enforcing an upper limit on $R$ also acts as a limit on the fraction of cells being reweighted. Figure \ref{fig:fracs} shows the increasing deviation from the original negatively weighted sample as $R$ increases, most visibly in \drjj.
\section{Metric choice}
A suitable cell metric groups kinematically similar events, which are likely to populate the same regions of observable distributions and therefore minimize bias from weight redistribution.

The original metric proposed in Refs. \cite{Andersen:2021mvw, Andersen:2024mqh} separated final-state objects into different categories and summed over modified Euclidean distances for each category:
\begin{equation*}
    D(p_j,p'_j)=\sqrt{|\vec{p}_j-\vec{p}_j'|^2+\tau^2(p_{T,j}-p'_{T,j})^2}.
    \label{eq:andersen}
\end{equation*}
The parameter $\tau$ controls the relative weighting between the spatial- and transverse-momentum terms, and was kept at the default value of $\tau=0$ for the following comparisons. If sets $t$ and $t'$ contain different particle multiplicities, a number of zero-momentum ``ghost'' particles are added until the two events have an equal number of particles.

To enforce infrared and collinear (IRC) safety, this method relies on clustering final-state hadrons into jets prior to reweighting. This introduces a dependence on the jet definition, including the choice of jet clustering algorithm and radius.

\subsection{Energy Mover's Distance}
The Energy Mover's Distance (EMD) \cite{Komiske:2019fks} is the particle physics counterpart to the Earth Mover's distance, which defines a distance metric between collider events by quantifying the minimum work needed to rearrange the radiation pattern or energy flow of one event into another. The EMD between two events $\mathcal{E}$ and $\mathcal{E}'$ is defined by
\begin{equation*}
    \mathrm{EMD}_{\beta,R}(\mathcal{E},\mathcal{E}')=\underset{[f_{ij}\geq0]}{\mathrm{min}}\sum_{i=1}^{N}\sum_{j=1}^{N'}f_{ij}\left(\frac{\theta_{ij}}{R}\right)^\beta + \left| \sum_{i=1}^NE_i -\sum_{j=1}^{N'}E_j'\right|,
    \label{eq:EMD}
\end{equation*}
where $f_{ij}$ is the transport plan specifying how much energy is moved from particle $i$ of $\mathcal{E}$ to particle $j$ of $\mathcal{E}'$, $E_i$ and $E_j'$ are the energies of the particles, and $\theta_{ij}$ is the ground distance between said particles. For proton-proton collisions at the LHC where general-purpose detectors are cylindrical about the beam axis, the ground metric is naturally parameterized by pseudorapidity $\eta$ and azimuthal angle $\phi$, and the particle energy is replaced by the transverse momentum $p_T$. 

\begin{figure}[!htbp]
\centering{
\includegraphics[width=0.32\textwidth]{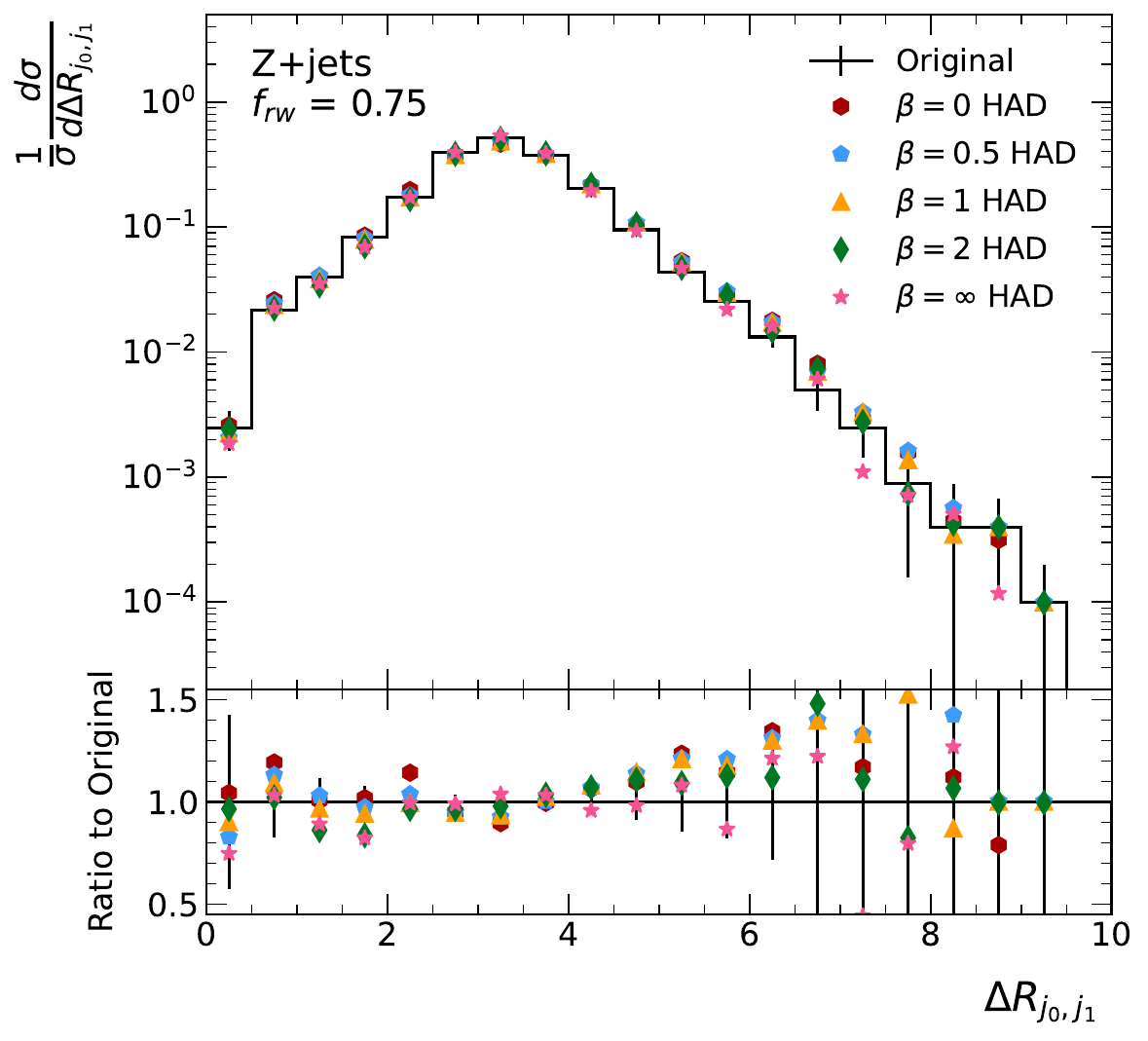}
\includegraphics[width=0.32\textwidth]{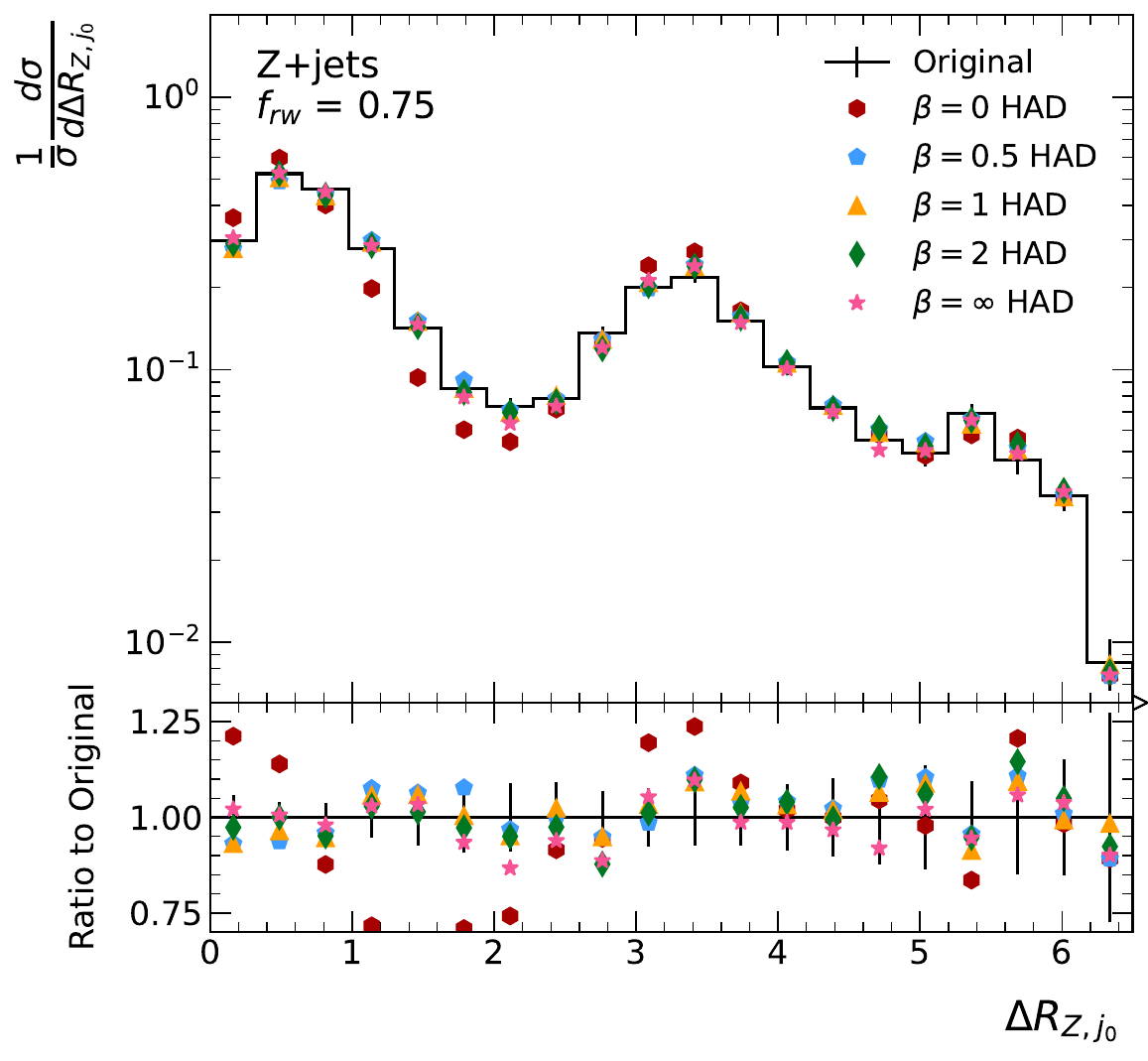}
\includegraphics[width=0.32\textwidth]{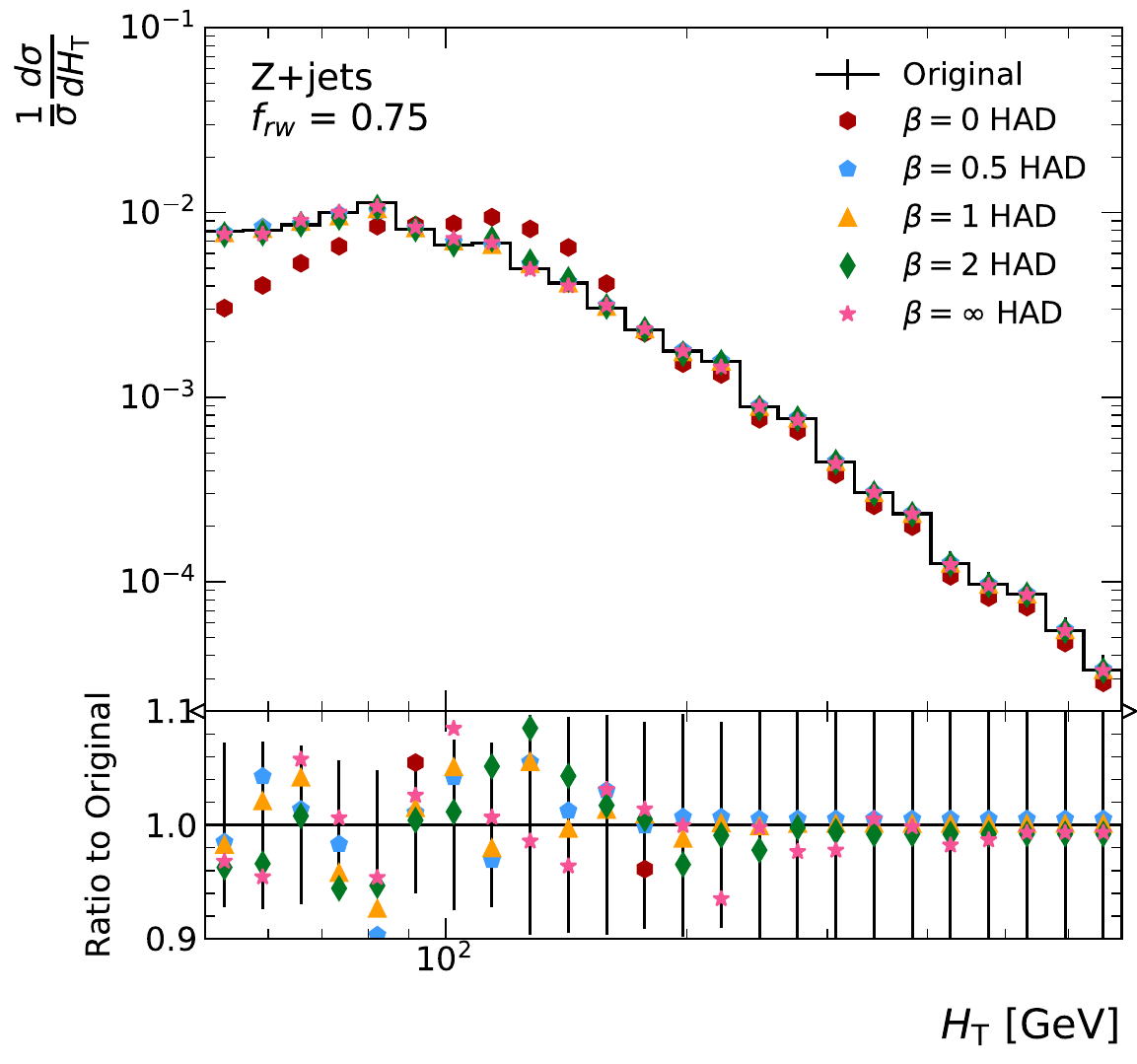}\\
 \includegraphics[width=0.32\textwidth]{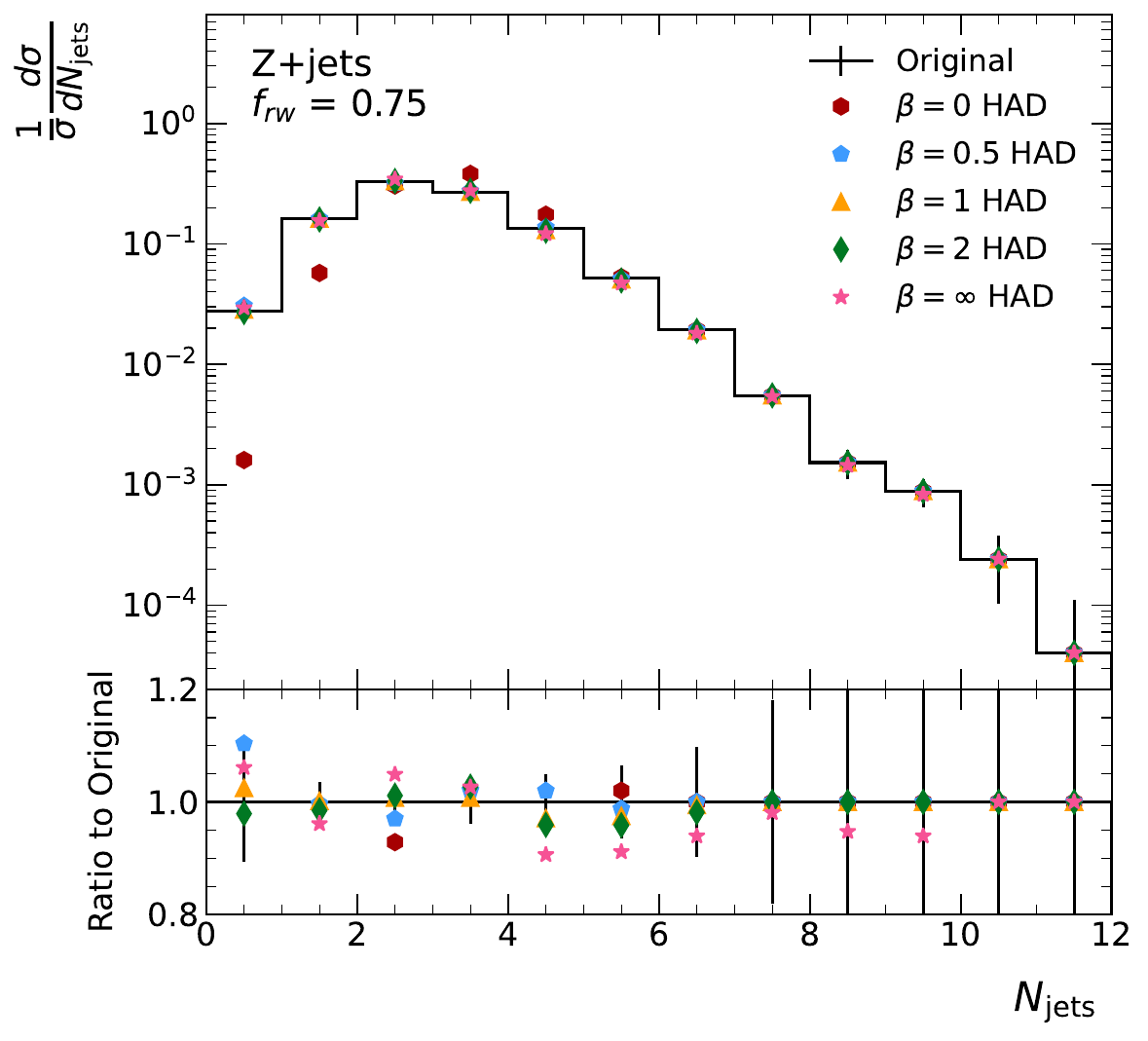}
\includegraphics[width=0.32\textwidth]{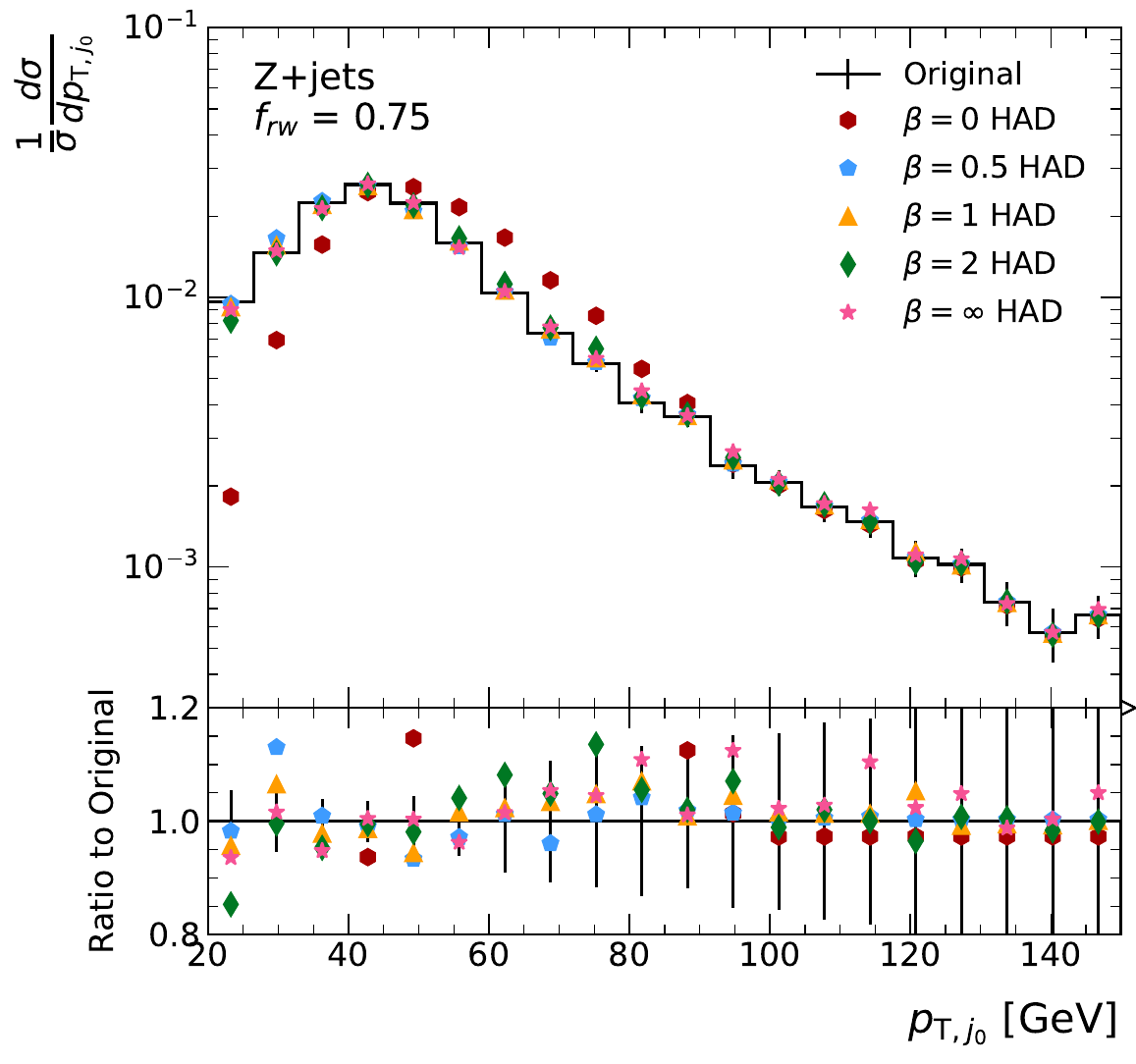}
\includegraphics[width=0.32\textwidth]{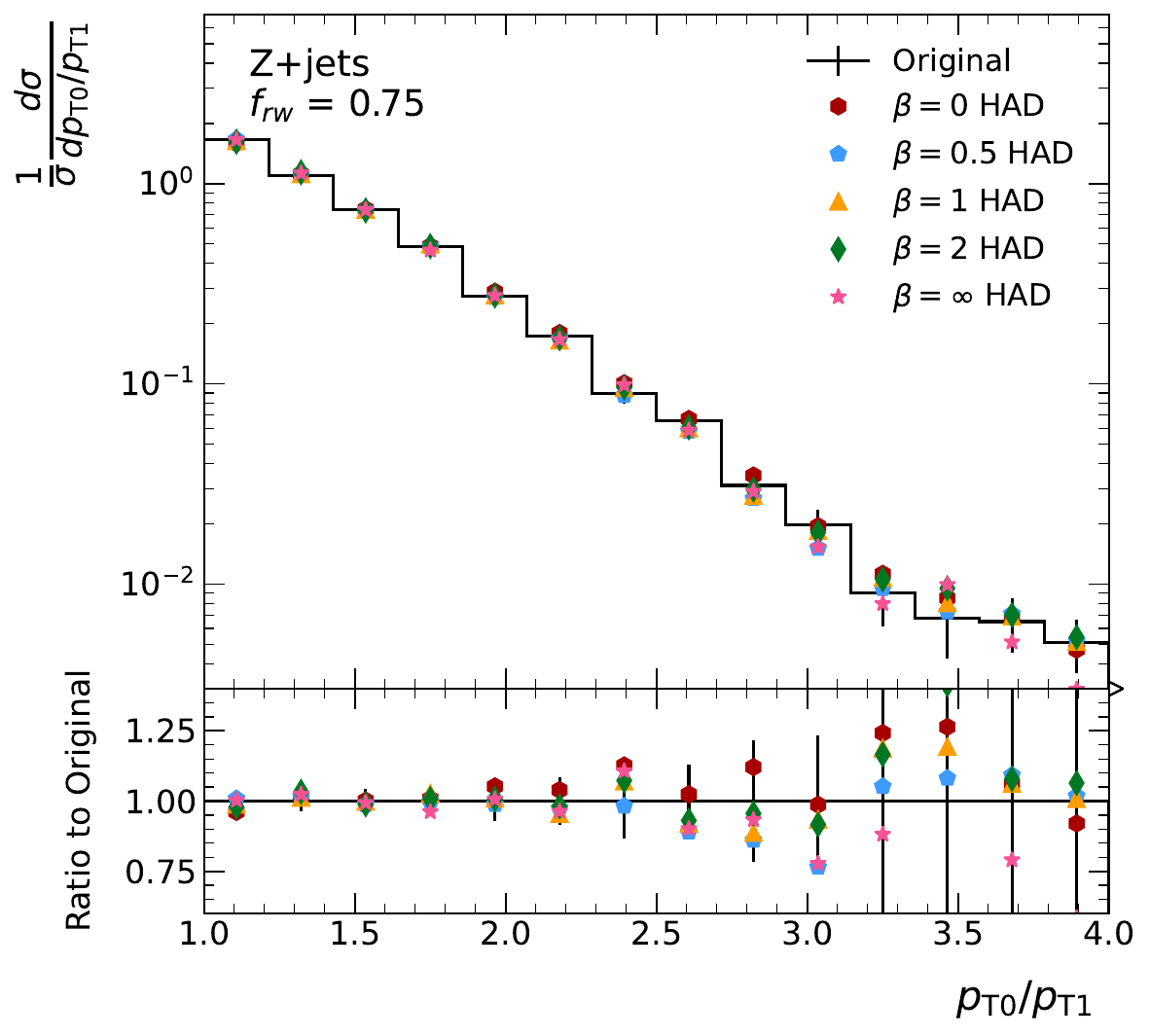}
    }
   \caption{Comparison of the normalized hadron-level unweighted \zjets observables to samples with varying $\beta$ values for the EMD, but fixed $f_{RW}=0.75$.}
 \label{fig:betas}
\end{figure}

The details of the implementation and calculation of the EMD can be found in Ref. \cite{Komiske:2019fks, Doherty:2026ot}, but we highlight the choice of the parameter $\beta$ here. This parameter controls the sensitivity of the metric to transport at different angular scales. We can see from Figure \ref{fig:betas} that across typical observables of \zjets events, $\beta=1$ and $\beta=0.5$ produce the smallest bias with respect to the original sample for a large reweight fraction. For all further studies, we use $\beta=1$ because the square root operation required for $\beta=0.5$ doubles the computational cost of the calculation.

\begin{figure}[!htbp]
\centering{
\includegraphics[width=0.32\textwidth]{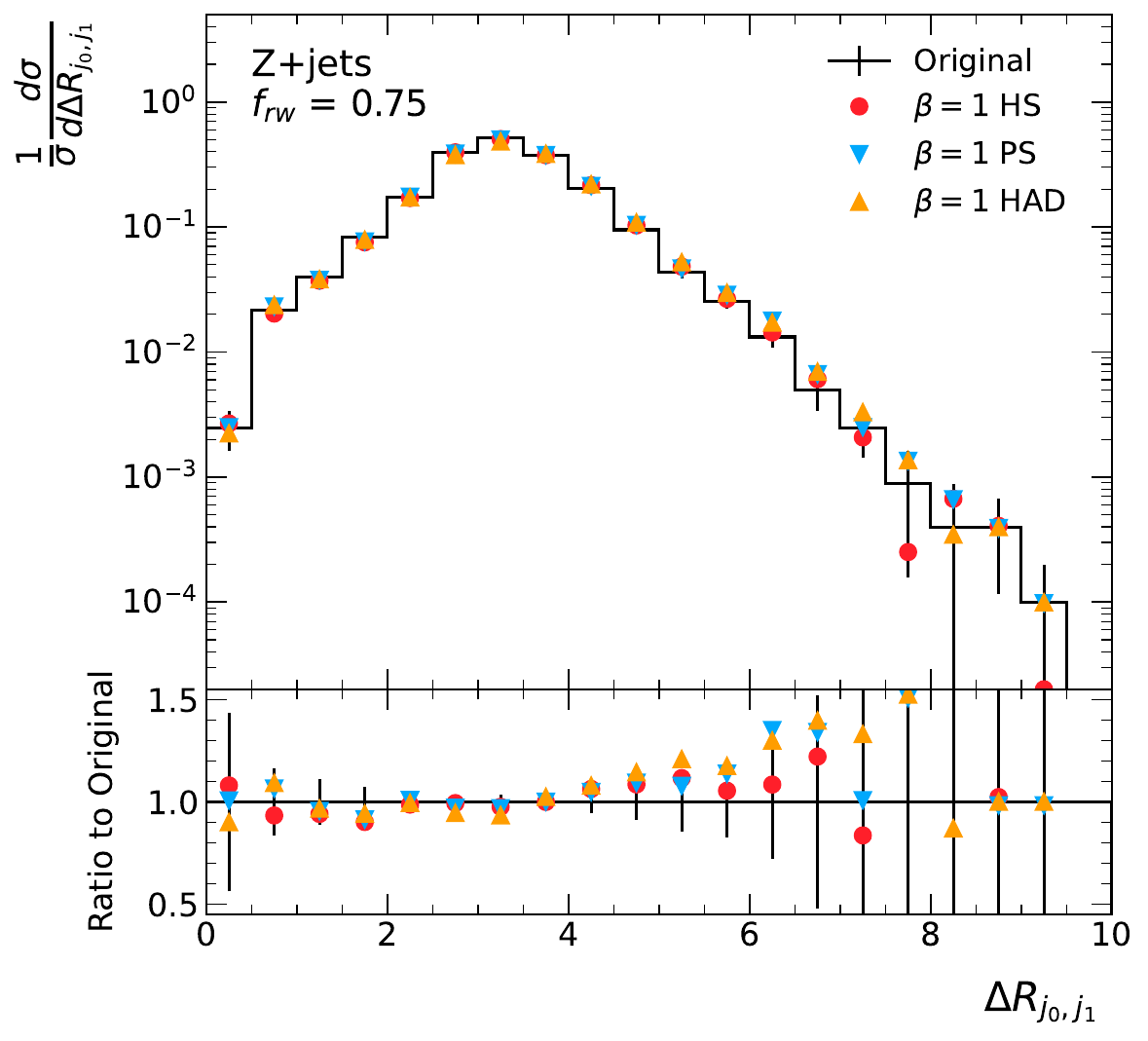}
\includegraphics[width=0.32\textwidth]{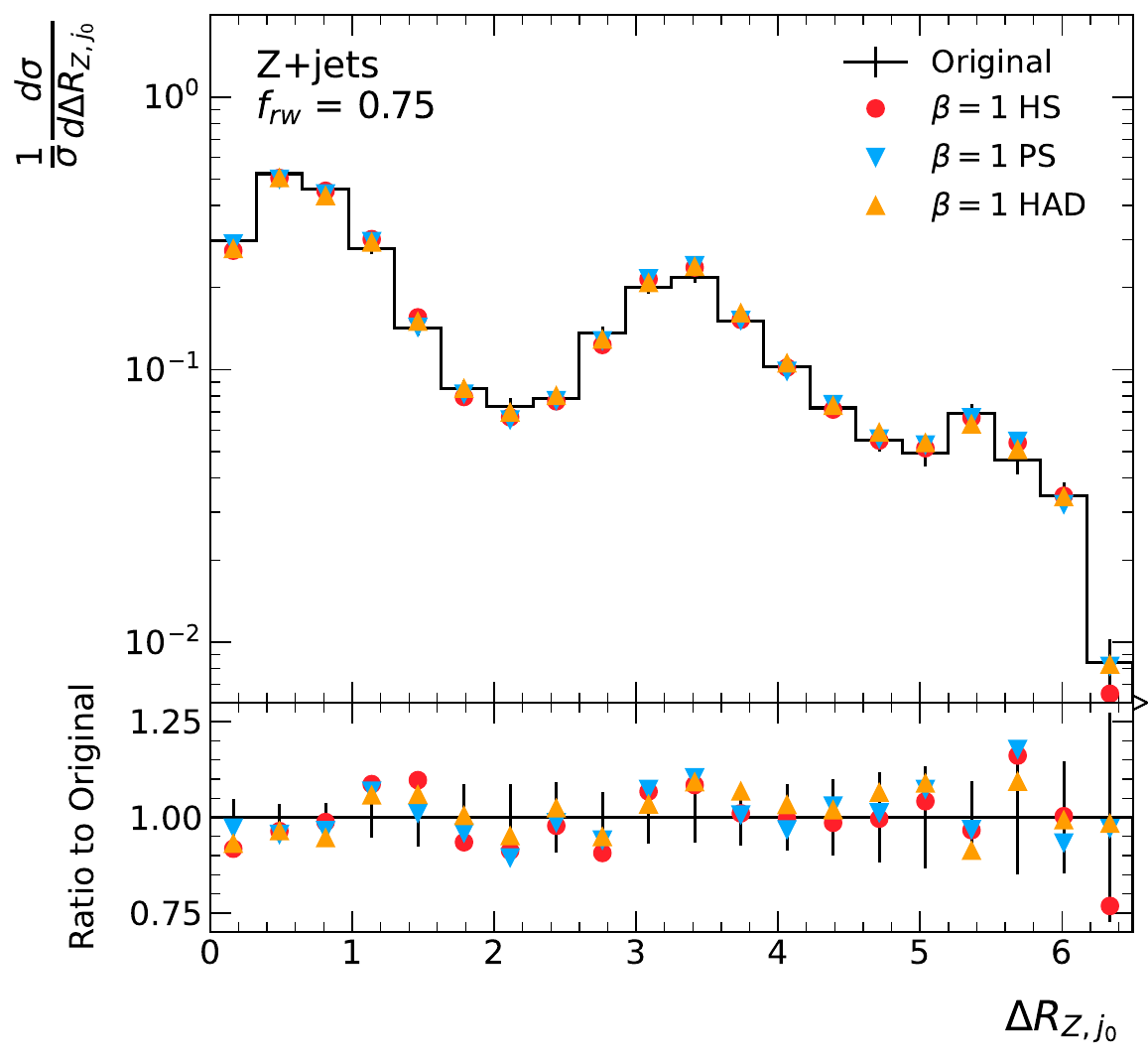}
\includegraphics[width=0.32\textwidth]{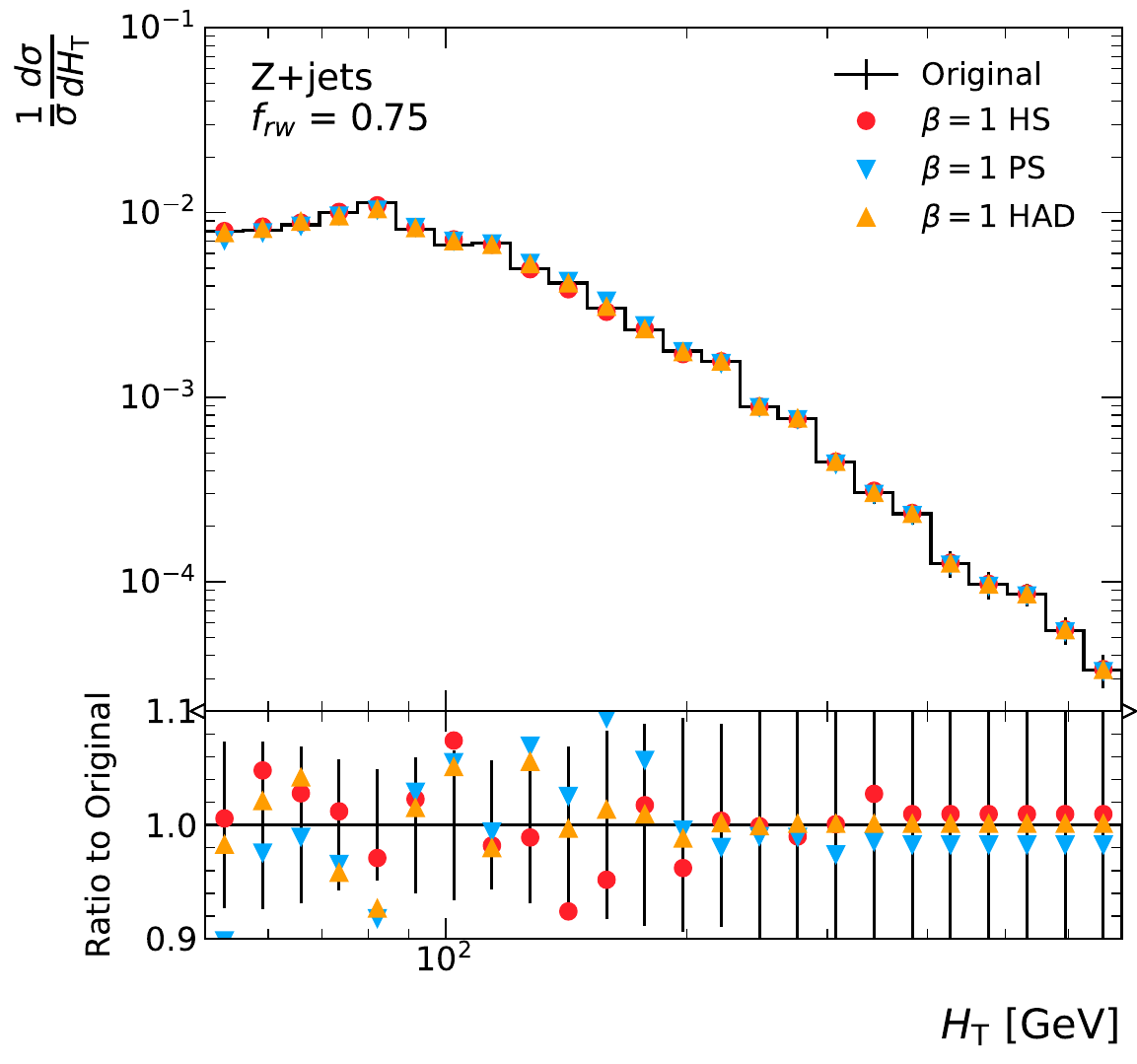} \\
\includegraphics[width=0.32\textwidth]{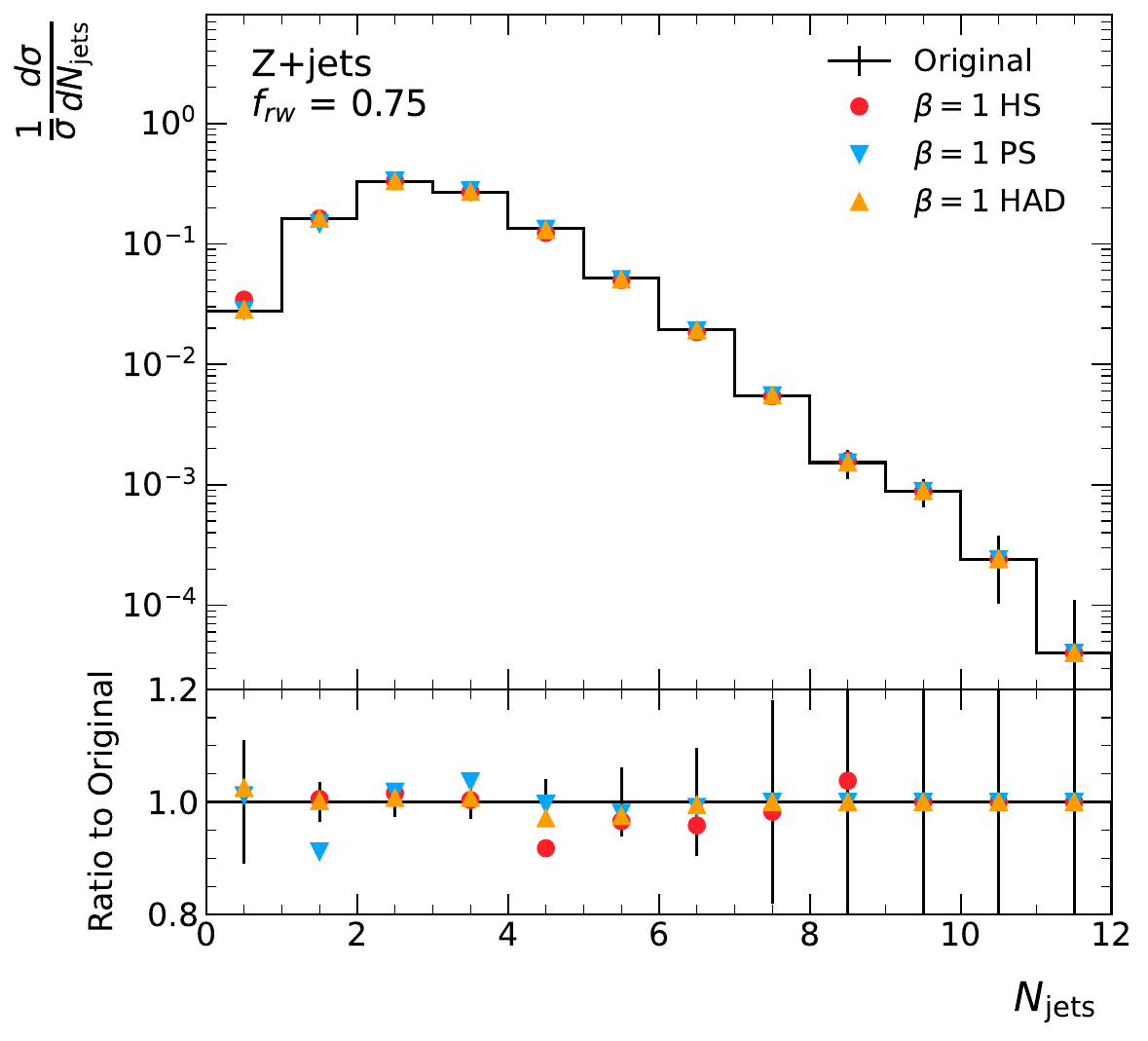}
   \includegraphics[width=0.32\textwidth]{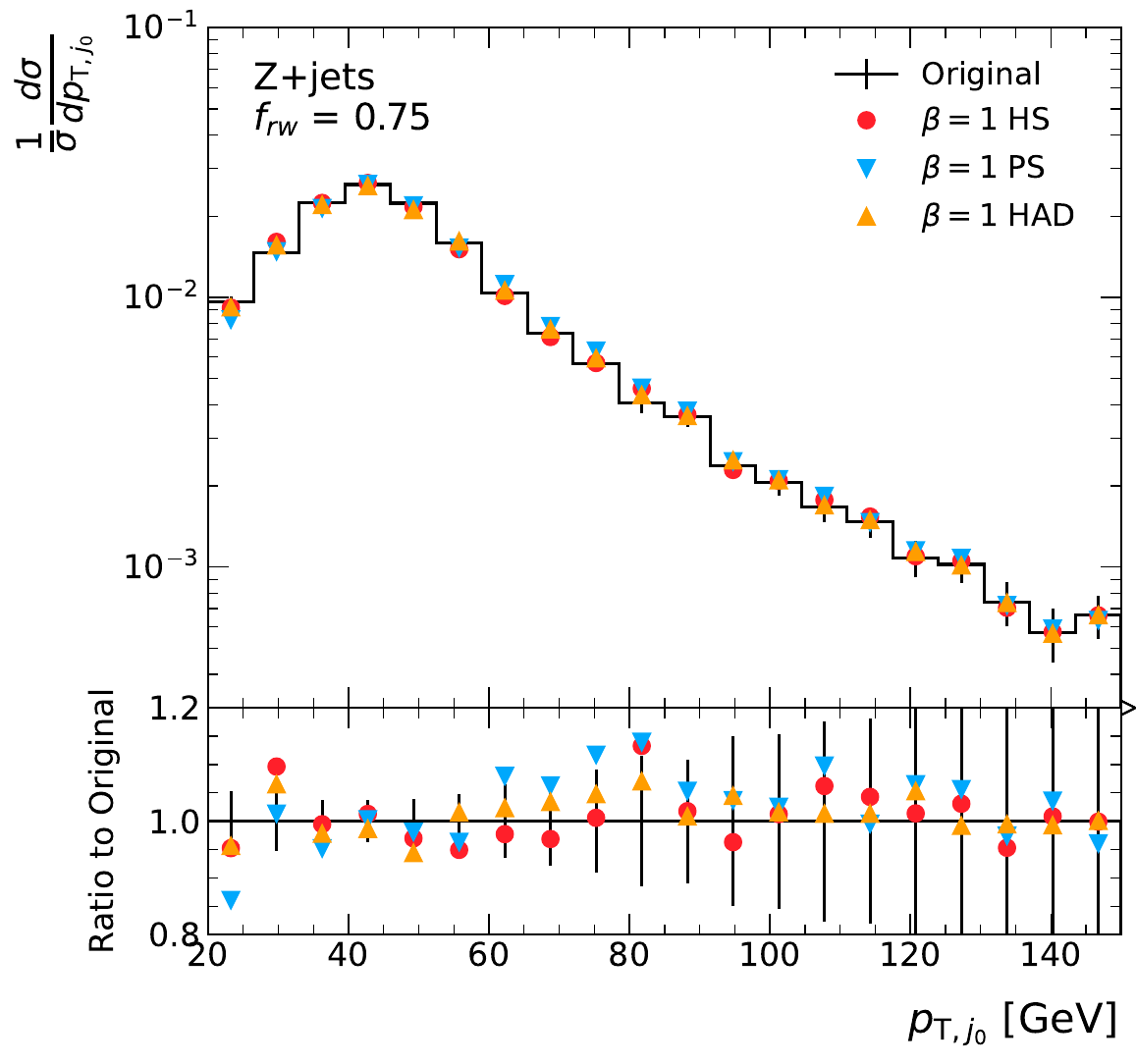}
\includegraphics[width=0.32\textwidth]{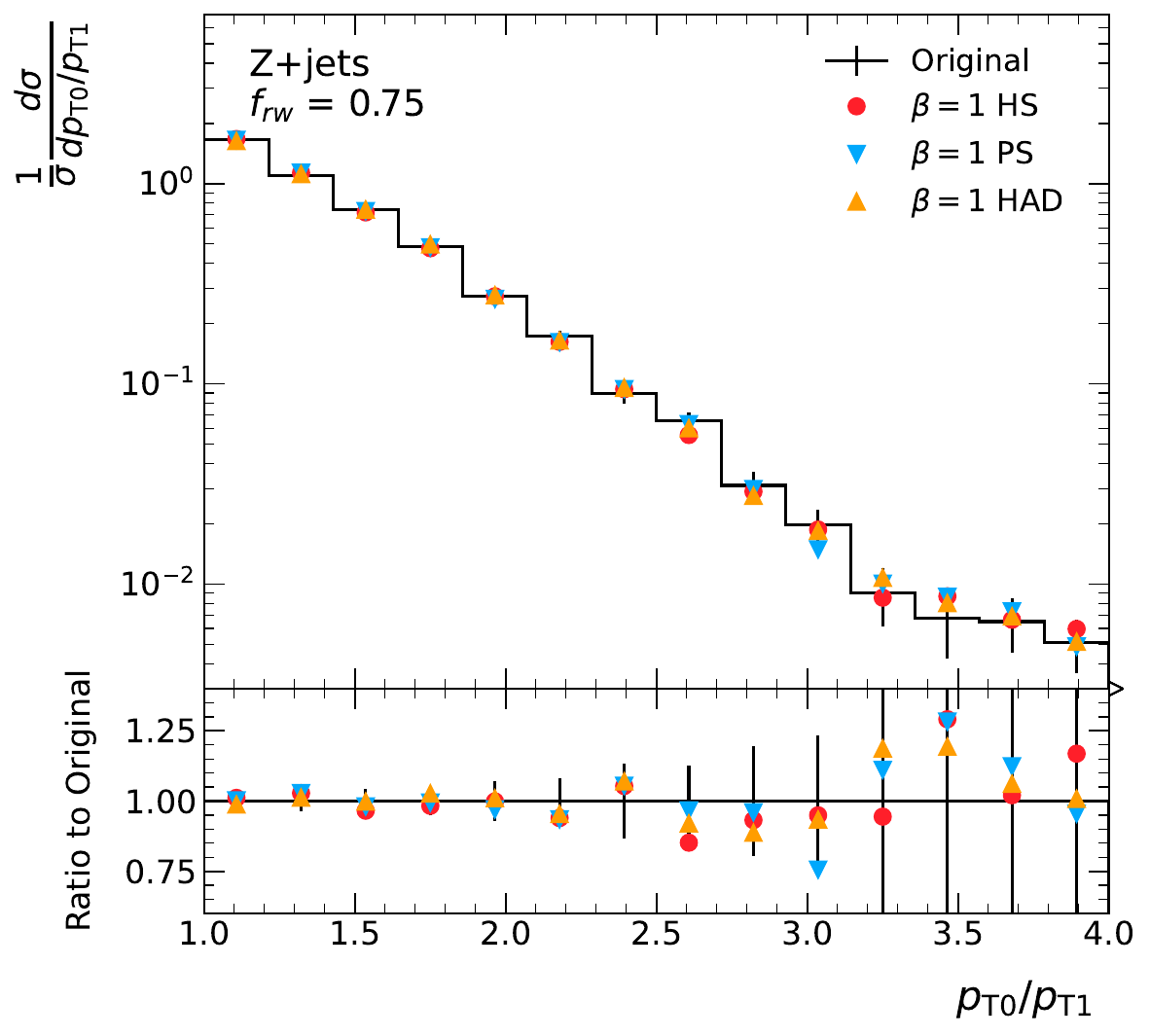}
    }
   \caption{Comparison of the normalized hadron-level unweighted \zjets observables to samples with the same final reweight fraction $f_{RW}=0.75$ using the EMD metric with $\beta=1$, evaluated at the different generation stages: HS, PS and HAD.}
 \label{fig:stages}
\end{figure}

Previous cell-reweighting studies have been performed exclusively on fully hadronized events to ensure that the procedure is IRC-safe \cite{Andersen:2021mvw}. Since the OT-based metrics we consider are themselves IRC-safe, they can be directly applied to events after any stage of event generation without intermediate jet clustering. We compare the performance of reweighting at three stages of event generation: at Born level after the hard-scattering matrix element calculation (HS), after the parton shower (PS), and after hadronization (HAD). We find that reweighting after hadronization biases the samples the least (Figure \ref{fig:stages}).
\subsection{Spectral Energy Mover's Distance}
Another OT-inspired metric is the Spectral Energy Mover's Distance (SEMD) \cite{Larkoski:2023qnv}. The SEMD operates on a one-dimensional representation of a single event's pairwise angular distances known as the \emph{spectral function}. For an event $\mathcal{E}$, this is defined as
\begin{equation*}
s(\omega)=\sum_{i,j\in\mathcal{E}}E_iE_j\delta(\omega-\omega_{ij}),
\end{equation*}
where $\omega_{ij}$ is the pairwise angular distance of particles $(i,j)$ and $E_i$ and $E_j$ are the individual particle energies.

\section{Evaluation}
To evaluate the different metric options, we generate proton-proton collision events at a center-of-mass energy of $\sqrt{s}=13$ TeV with a final state of a $Z$ boson produced in association with jets (\zjets). The matrix elements are calculated at next-to-leading order (NLO) using \textsc{MadGraph5}\_a{MC}@{NLO} \cite{Alwall:2014hca}. Parton showering, hadronization, multi-parton interactions and the underlying event are modeled with \textsc{Pythia 8.3} \cite{Bierlich:2022pfr}. Following the complete event generation procedure, the resulting \zjets sample has a negative weight fraction of 37.6\%.

At the analysis level, final-state particles are required to have $p_T > 0.1$ GeV and $|\eta|< 4.9$. Jets are reconstructed using the anti-$k_t$ algorithm \cite{Cacciari:2008gp} with a radius parameter of $R = 0.4$, as implemented with the FastJet software package \cite{Fastjet}. Selected jets are required to have $p_T > 20$ GeV and to be within $|\eta|<4.5$. It is after these selections and jet reconstruction that we construct the observables shown in Figures \ref{fig:betas} and \ref{fig:stages}.

\begin{figure}[!htbp]
\centering{
\includegraphics[width=0.32\textwidth]{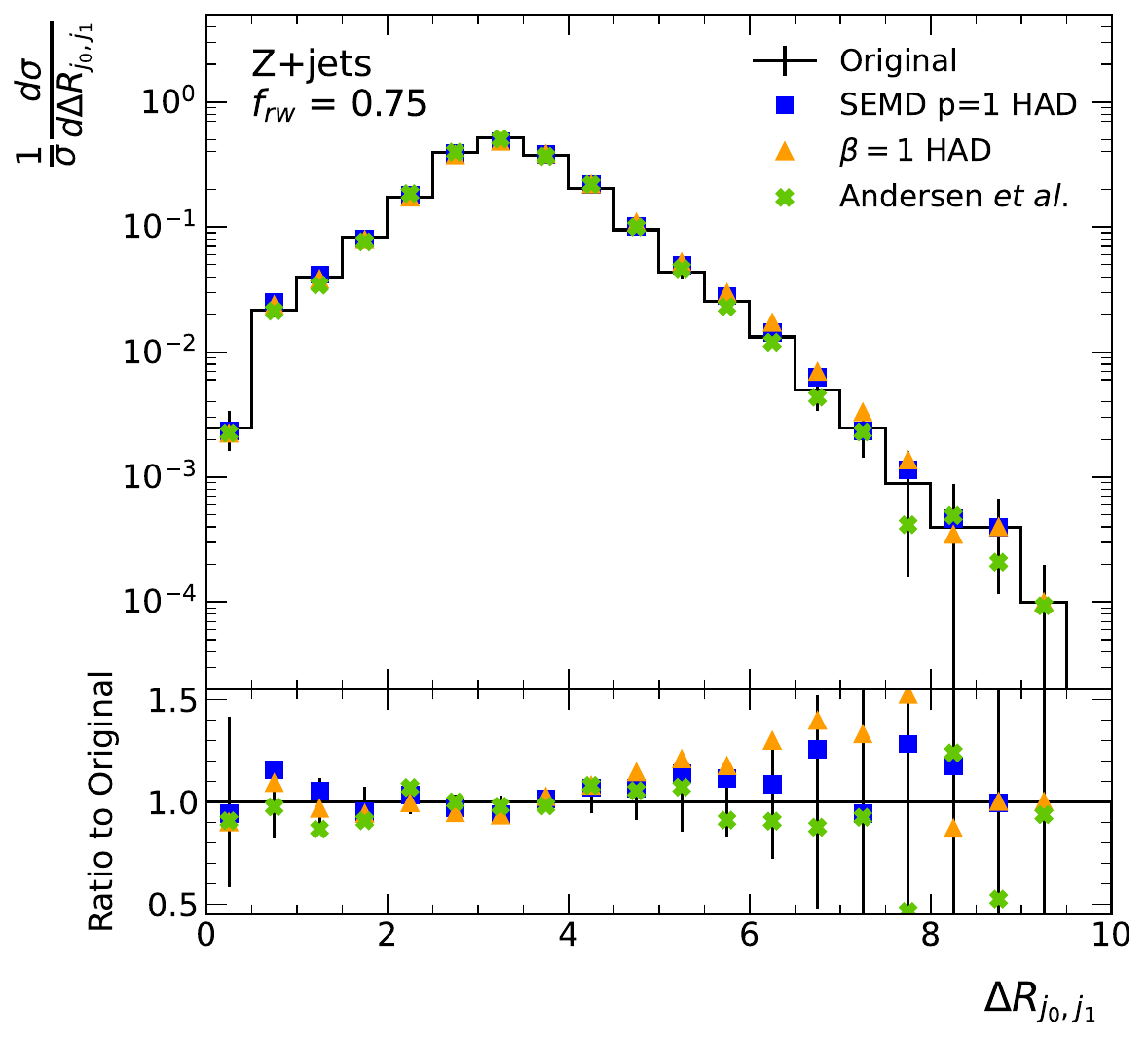}
\includegraphics[width=0.32\textwidth]{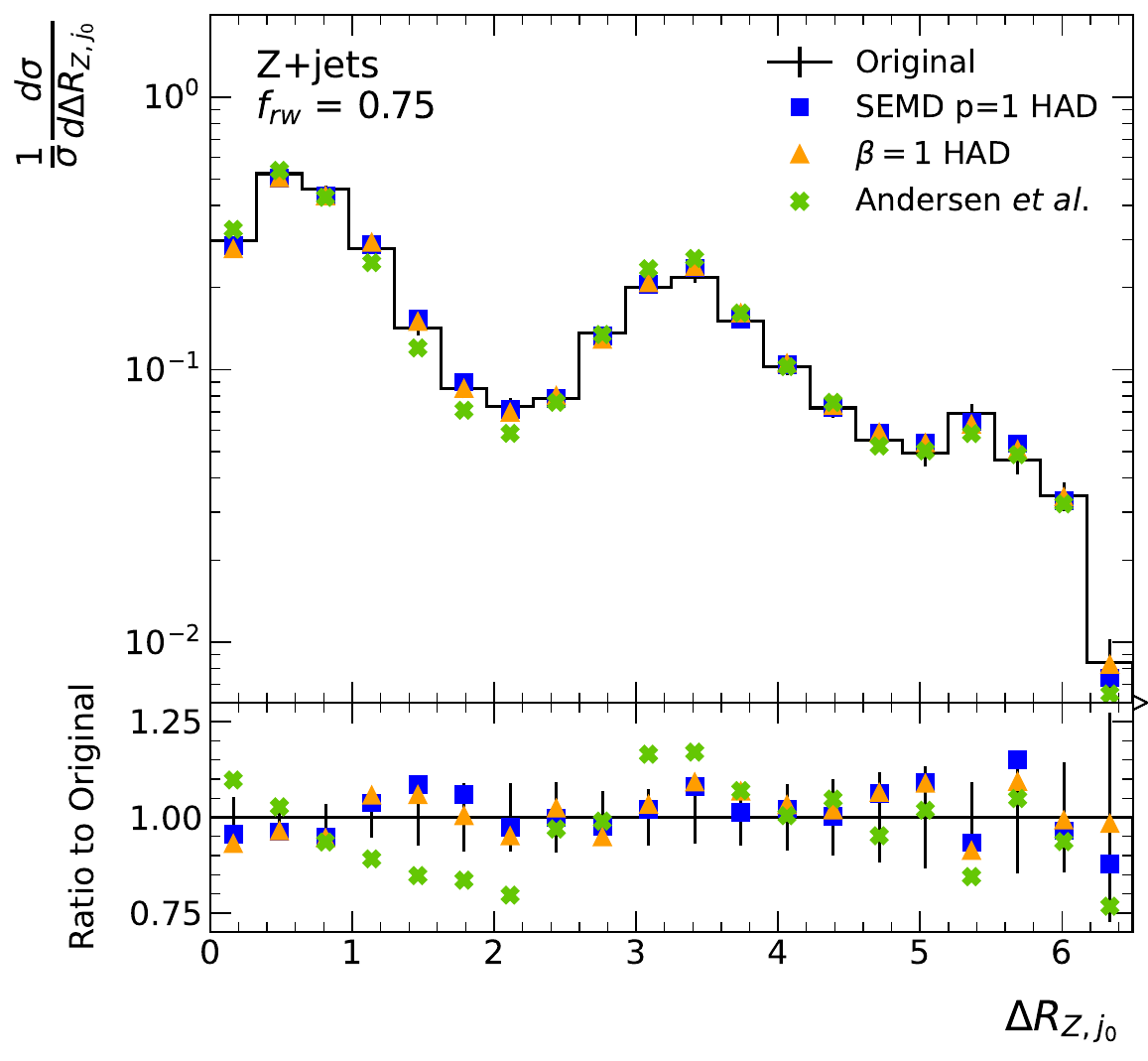}
\includegraphics[width=0.32\textwidth]{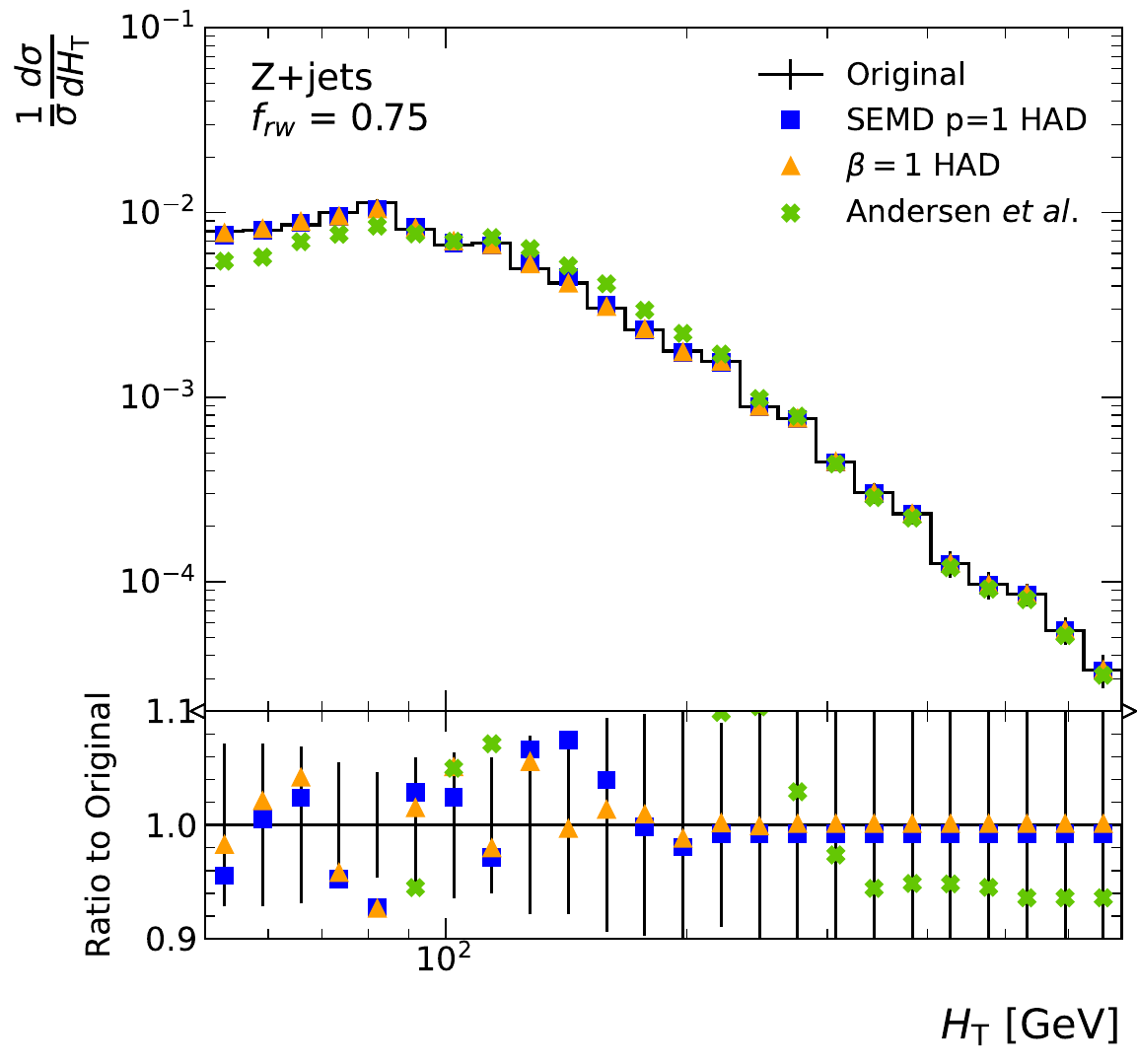} \\
\includegraphics[width=0.32\textwidth]{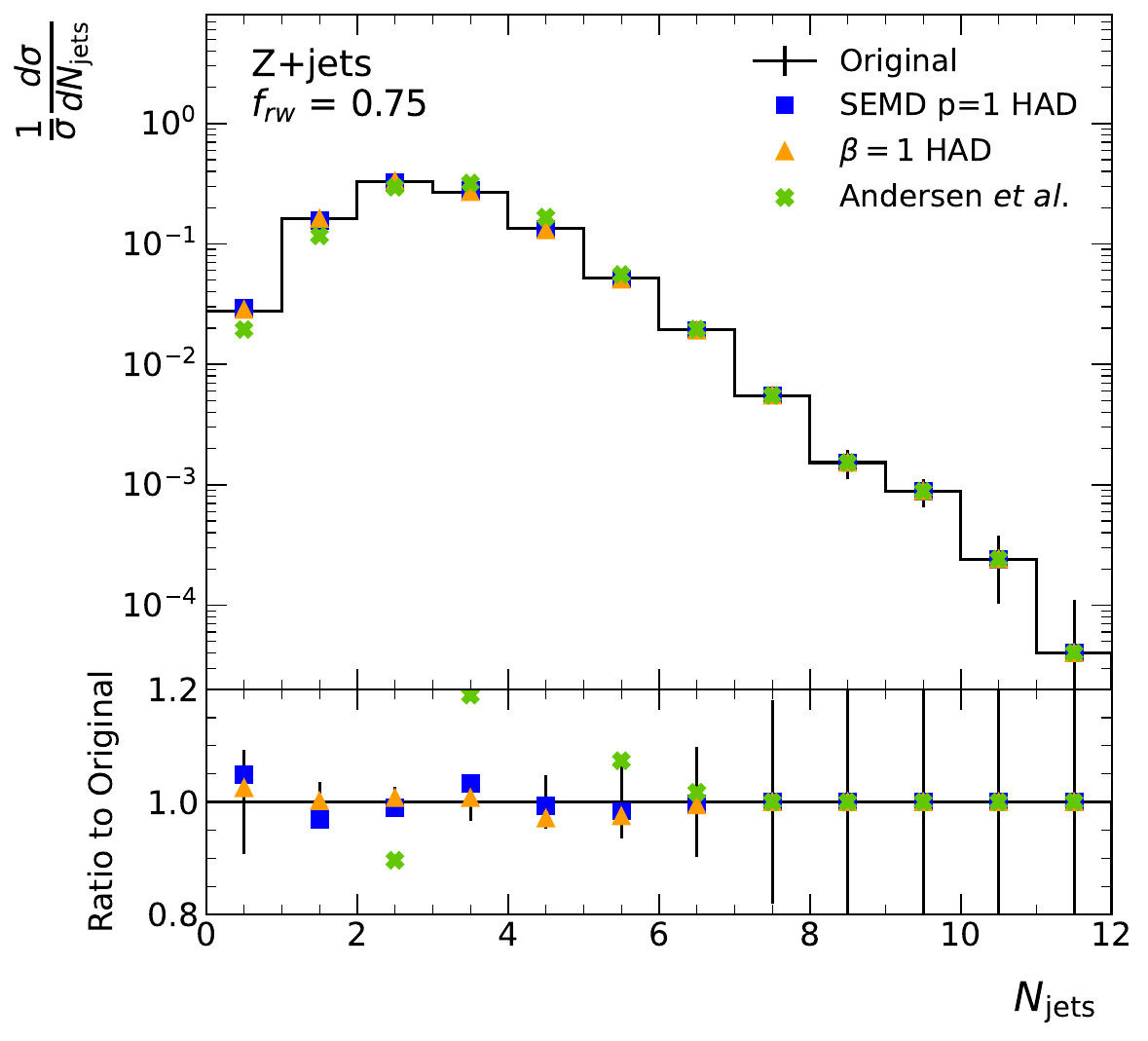}
\includegraphics[width=0.32\textwidth]{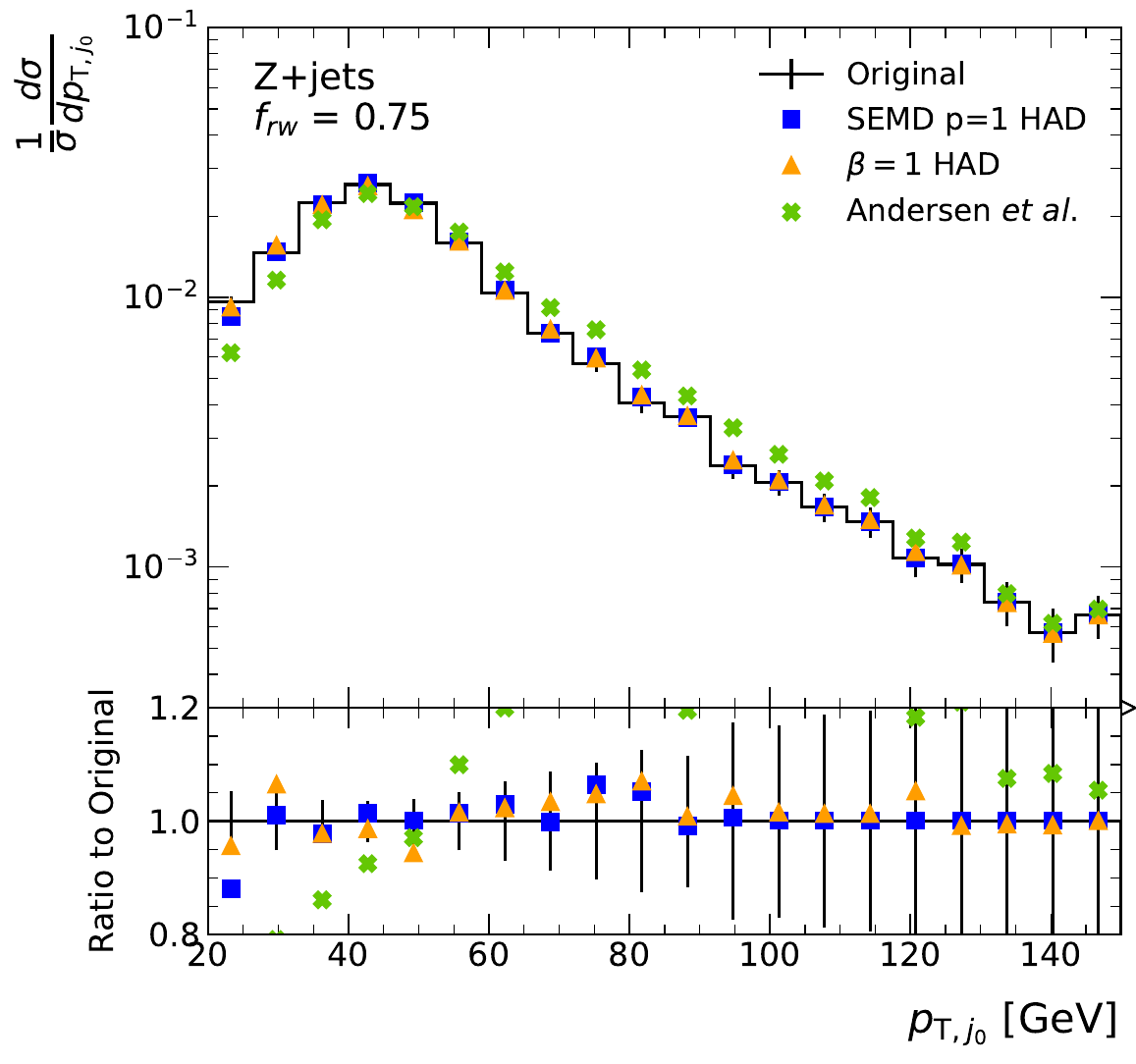}
\includegraphics[width=0.32\textwidth]{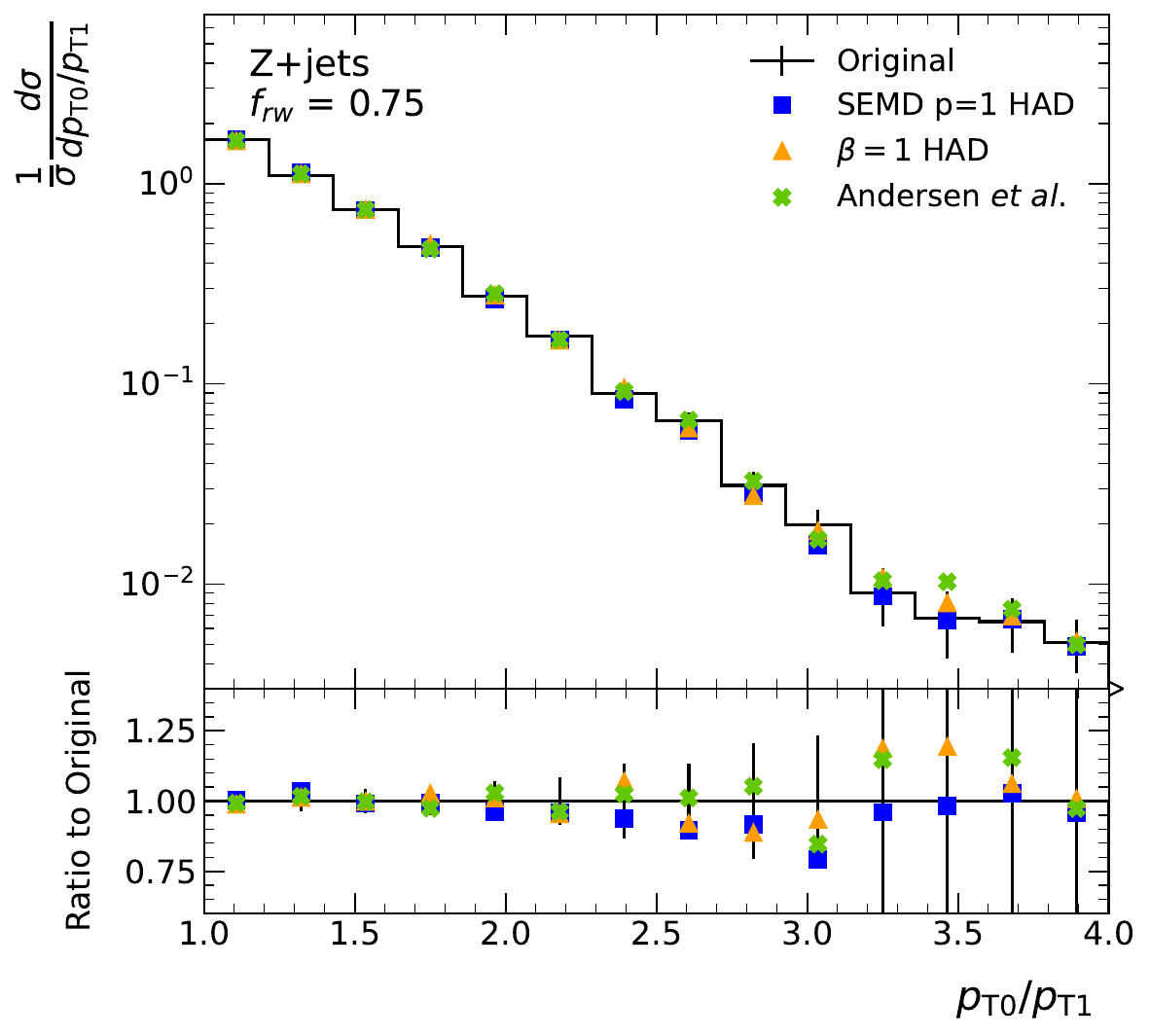}
   \caption{Comparison of the normalized hadron-level unweighted Z+jets observables to samples with the same final $f_{RW}=0.75$ using the three considered metrics.}
 \label{fig:metric}
 }
\end{figure}

A final evaluation of the original distance metric, SEMD, and EMD at the hadronization level can be seen in Figure \ref{fig:metric}. We see that the OT-inspired metrics perform better than Andersen \emph{et al.}'s metric, especially in \Ht and leading jet $p_T$.
For cases like this, where the differences between samples are difficult to evaluate via histograms alone, we employ an additional higher-dimensional metric known as the Theory Movers Distance (\XMD) \cite{Komiske:2020qhg}. Like the EMD, the \XMD measures the minimum ``work'' required to rearrange one set into another, but encodes the cost of rearranging theories weighted by their cross sections rather than event energy flow. A smaller \XMD between the reweighted sample and the original sample indicates that less bias has been introduced by the reweighting.

\begin{figure}[!htbp]
\centering{
   \subfloat[]{\includegraphics[width=0.33\textwidth]{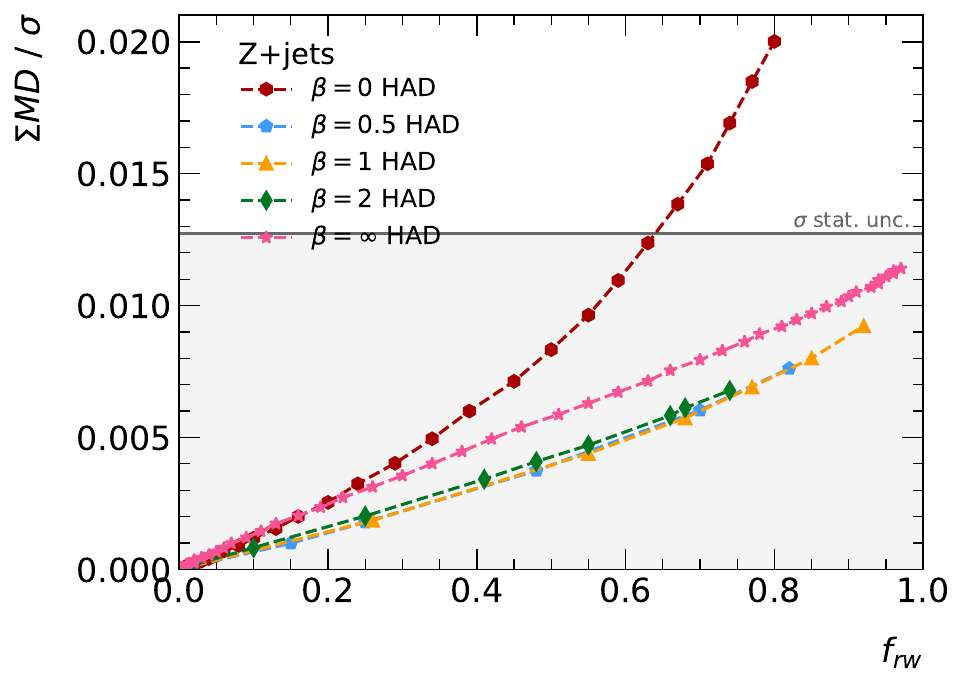}\label{fig:betas:xmd}}
   \subfloat[]{\includegraphics[width=0.33\textwidth]{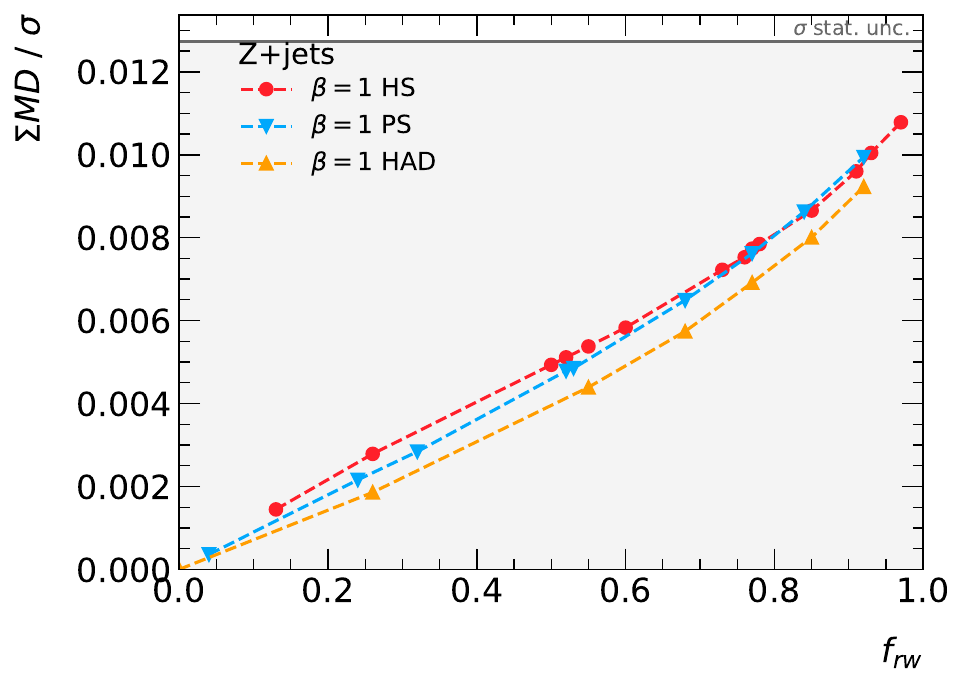}\label{fig:stages:xmd}}
    \subfloat[]{\includegraphics[width=0.33\textwidth]{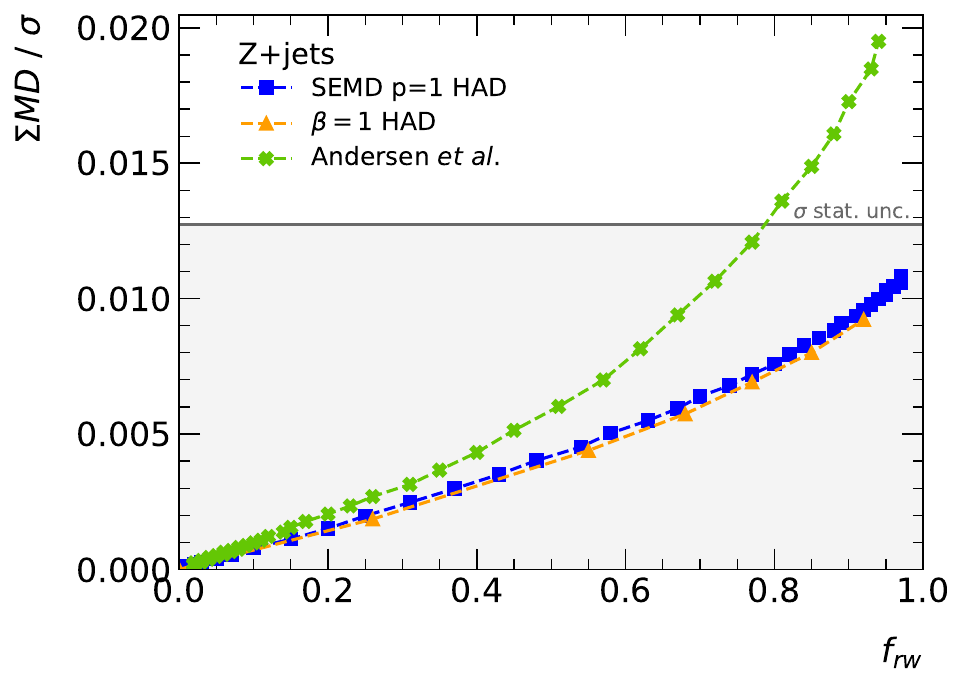}\label{fig:semdvemd:xmd}}
   \caption{A comparison of the \XMD vs $f_{RW}$ for the different distance metrics considered in this study. The grey line is the statistical uncertainty of the \zjets sample and the shaded region indicates which distances are within this uncertainty.}
 \label{fig:xmds}
 }
\end{figure}

\begin{figure}[!htbp]
\centering{
\sidecaption{\includegraphics[width=0.45\textwidth]{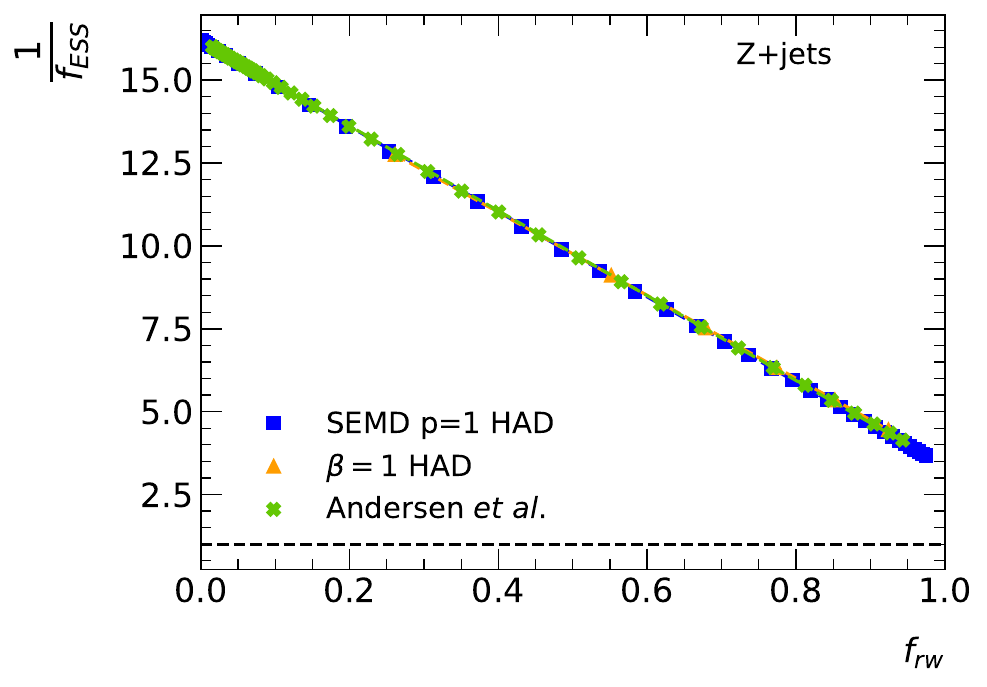}\label{fig:metric:kish_eff}}
   \caption{The value of $1/f_{ESS}$ as a function of reweighted fraction $f_{RW}$ for the different distance metrics.}
 \label{fig:dilution} }
\end{figure}

Looking at the \XMDs for the three metric options (Figure \ref{fig:semdvemd:xmd}), it is again apparent that the SEMD and EMD perform comparably, with the EMD performing slightly better. We can also see from Figures \ref{fig:betas} and \ref{fig:stages} that the \XMD points to the same conclusions for choice of $\beta$ and stage of generation as the one-dimensional observables.
Although cell-resampling may improve the statistical power of a sample by removing negative weights, the variance of the new nonuniform weights offsets part of this gain. We quantify the net improvement using the Kish effective sample fraction $f_{ESS}$ \cite{Kish1992}, as proposed in Ref. \cite{Frederix:2020trv}:
\begin{equation}
    f_{ESS}=\frac{1}{N}\frac{(\sum_iw_i)^2}{\sum_iw_i^2},
\end{equation}
where $w_i$ is the weight of event $i$ and $N$ is the total number of events. The Kish effective sample fraction quantifies the statistical power of a sample relative to a fully unweighted sample of equal size. Its inverse, $1/f_{ESS}$, gives the factor by which the weighted sample size must be increased to match the statistical power of the unweighted sample. We can see in Figure \ref{fig:dilution} that, at a reweight fraction $f_{RW}=0.75$, all cell-resampling variants more than double the effective statistical power, reducing $1/f_{ESS}$ from 16 to 6.
\vspace{-1.0\baselineskip}
\section{Conclusion}
Cell-resampling is an effective method to reduce negatively weighted event fractions in MC simulations, addressing one of the pressures on computing resources that will be faced in the HL-LHC era. In these studies, we have shown that using an optimal-transport-based metric to group events within cells introduces less bias compared to the Euclidean distance metric used in earlier implementations.

We intentionally choose a high $f_{RW}=0.75$ to demonstrate the extremes of the bias that can be incurred in a low-statistics sample with this method. This bias would decrease as statistics are increased, but even in the conditions evaluated in this study the \XMD between the negatively weighted sample and the OT-based cell-resampled samples is less than the statistical uncertainty of the MC sample, $\delta\sigma/\sigma = \sqrt{\sum_i w_i^2}/\sum_i w_i = 0.01273$.

After studying the performance of EMD reweighting with various choices of the angular parameter, $\beta=1$ was found to give the best combination of accuracy and computational efficiency. The IRC-safety of the OT-based metrics means that the resampling can be performed at any stage of event generation; we found that reweighting after hadronization gave the most stable performance across all observables and the \XMD. Both OT-based metrics performed comparably well for cell-resampling and provided a substantial improvement over the modified-Euclidean distance originally proposed for the method.

Since this method is process-independent, it could be applied in conjunction with other negative weight mitigation techniques. To support reproducibility, the code and datasets used in this study are publicly available in Refs. \cite{doherty_2026_21284997,otcres_v1.0.0}.
\section*{Acknowledgements}
This material is based on work supported by the U.S. Department of Energy, Office of Science, Office of High Energy Physics under Award No.~DE-SC0026285. This work is also supported by the National Science Foundation under Cooperative Agreement PHY-2019786 (NSF AI Institute for Artificial Intelligence \& Fundamental Interactions, \url{http://iaifi.org/}). We also acknowledge early support from a seed grant from the Brown University Data Science Institute.
\bibliography{ref}

@dataset{doherty_2026_21284997,
  author       = {Doherty, Regan and
                  Hay, Lauren and
                  Jain, Rishabh and
                  LeBlanc, Matt and
                  Marrinan, Julia and
                  Mauceri, Camille and
                  Roloff, Jennifer},
  title        = {MadGraph5\_aMC@NLO + Pythia 8.3 event samples for
                   cell resampling studiess
                  },
  month        = jul,
  year         = 2026,
  publisher    = {Zenodo},
  version      = {1.0.0},
  doi          = {10.5281/zenodo.21284997},
  url          = {https://doi.org/10.5281/zenodo.21284997},
}

@software{otcres_v1.0.0,
  author    = {Doherty, Regan and Hay, Lauren and Jain, Rishabh and LeBlanc, Matt and Marrinan, Julia and Mauceri, Camille and Roloff, Jennifer},
  title     = {ot-cres: optimal-transport-based cell resampling for negative event weights},
  version   = {v1.0.0},
  publisher = {Zenodo},
  year      = {2026},
  doi       = {10.5281/zenodo.22711649},
  url       = {https://github.com/leblanc-lab/ot-cres}
}

@article{Komiske:2019fks,
    author = "Komiske, Patrick T. and Metodiev, Eric M. and Thaler, Jesse",
    title = "{Metric Space of Collider Events}",
    eprint = "1902.02346",
    archivePrefix = "arXiv",
    primaryClass = "hep-ph",
    reportNumber = "MIT-CTP 5102",
    doi = "10.1103/PhysRevLett.123.041801",
    journal = "Phys. Rev. Lett.",
    volume = "123",
    number = "4",
    pages = "041801",
    year = "2019"
}

@article{Komiske:2020qhg,
    author = "Komiske, Patrick T. and Metodiev, Eric M. and Thaler, Jesse",
    title = "{The Hidden Geometry of Particle Collisions}",
    eprint = "2004.04159",
    archivePrefix = "arXiv",
    primaryClass = "hep-ph",
    reportNumber = "MIT-CTP 5185",
    doi = "10.1007/JHEP07(2020)006",
    journal = "JHEP",
    volume = "07",
    pages = "006",
    year = "2020"
}

@article{Kish1992,
  author  = {Kish, Leslie},
  title   = {Weighting for Unequal $P_i$},
  journal = {Journal of Official Statistics},
  volume  = {8},
  number  = {2},
  pages   = {183--200},
  year    = {1992}
}

@article{Frederix:2020trv,
    author = "Frederix, R. and Frixione, S. and Prestel, S. and Torrielli, P.",
    title = "{On the reduction of negative weights in MC@NLO-type matching procedures}",
    eprint = "2002.12716",
    archivePrefix = "arXiv",
    primaryClass = "hep-ph",
    reportNumber = "LU-TP 20-09, MCNET-20-08",
    doi = "10.1007/JHEP07(2020)238",
    journal = "JHEP",
    volume = "07",
    pages = "238",
    year = "2020"
}

@Article{Fastjet,
      author         = "Cacciari, Matteo and Salam, Gavin P. and Soyez, Gregory",
      title          = "{FastJet user manual}",
      journal        = "Eur. Phys. J. C",
      volume         = "72",
      year           = "2012",
      pages          = "1896",
      doi            = "10.1140/epjc/s10052-012-1896-2",
      eprint         = "1111.6097",
      archivePrefix  = "arXiv",
      primaryClass   = "hep-ph",
      reportNumber   = "CERN-PH-TH-2011-297",
      SLACcitation   = "%%CITATION = ARXIV:1111.6097;%%"
}

@Article{Cacciari:2008gp,
     author    = "Cacciari, Matteo and Salam, Gavin P. and Soyez, Gregory",
     title     = "{The anti-\(k_{t}\) jet clustering algorithm}",
     journal   = "JHEP",
     volume    = "04",
     year      = "2008",
     pages     = "063",
     eprint    = "0802.1189",
     archivePrefix = "arXiv",
     primaryClass  =  "hep-ph",
     doi       = "10.1088/1126-6708/2008/04/063",
     SLACcitation  = "%%CITATION = 0802.1189;%%"
}

@article{Bierlich:2022pfr,
    author = "Bierlich, Christian and others",
    title = "{A comprehensive guide to the physics and usage of PYTHIA 8.3}",
    eprint = "2203.11601",
    archivePrefix = "arXiv",
    primaryClass = "hep-ph",
    reportNumber = "LU-TP 22-16, MCNET-22-04, FERMILAB-PUB-22-227-SCD",
    doi = "10.21468/SciPostPhysCodeb.8",
    journal = "SciPost Phys. Codeb.",
    volume = "2022",
    pages = "8",
    year = "2022"
}

@Article{Alwall:2014hca,
      author         = "Alwall, J. and Frederix, R. and Frixione, S. and Hirschi,
                        V. and Maltoni, F. and Mattelaer, O. and Shao, H. -S. and
                        Stelzer, T. and Torrielli, P. and Zaro, M.",
      title          = "{The automated computation of tree-level and
                        next-to-leading order differential cross sections, and
                        their matching to parton shower simulations}",
      journal        = "JHEP",
      volume         = "07",
      year           = "2014",
      pages          = "079",
      doi            = "10.1007/JHEP07(2014)079",
      eprint         = "1405.0301",
      archivePrefix  = "arXiv",
      primaryClass   = "hep-ph",
      reportNumber   = "CERN-PH-TH-2014-064, CP3-14-18, LPN14-066, MCNET-14-09,
                        ZU-TH-14-14",
      SLACcitation   = "%%CITATION = ARXIV:1405.0301;%%"
}

@article{Larkoski:2023qnv,
    author = "Larkoski, Andrew J. and Thaler, Jesse",
    title = "{A spectral metric for collider geometry}",
    eprint = "2305.03751",
    archivePrefix = "arXiv",
    primaryClass = "hep-ph",
    reportNumber = "MIT-CTP 5556",
    doi = "10.1007/JHEP08(2023)107",
    journal = "JHEP",
    volume = "08",
    pages = "107",
    year = "2023"
}

@misc{Doherty:2026ot,
      title={Optimal-Transport-Based Cell Resampling for Negative and Pathological Event Weights}, 
      author={Regan Doherty and Lauren Hay and Rishabh Jain and Matt LeBlanc and Julia Marrinan and Camille Mauceri and Jennifer Roloff},
      year={2026},
      eprint={2607.08723},
      archivePrefix={arXiv},
      primaryClass={hep-ph},
      url={https://arxiv.org/abs/2607.08723}, 
}

@techreport{ATLAS_CDR,
      author        = "{ATLAS Collaboration}",
      title         = "{ATLAS Software and Computing HL-LHC Roadmap}",
      institution   = "CERN",
      reportNumber  = "CERN-LHCC-2022-005",
      address       = "Geneva",
      year          = "2022",
      url           = "https://cds.cern.ch/record/2802918",
}

@article{Andersen:2021mvw,
    author = "Andersen, Jeppe R. and Maier, Andreas",
    title = "{Unbiased elimination of negative weights in Monte Carlo samples}",
    eprint = "2109.07851",
    archivePrefix = "arXiv",
    primaryClass = "hep-ph",
    reportNumber = "DCPT/21/54, DESY 21-135, IPPP/21/27, MCNET-21-14, SAGEX-21-29",
    doi = "10.1140/epjc/s10052-022-10372-3",
    journal = "Eur. Phys. J. C",
    volume = "82",
    number = "5",
    pages = "433",
    year = "2022"
}

@article{Andersen:2024mqh,
    author = "Andersen, Jeppe R. and Cueto, Ana and Jones, Stephen P. and Maier, Andreas",
    title = "{A Cell Resampler study of Negative Weights in Multi-jet Merged Samples}",
    eprint = "2411.11651",
    archivePrefix = "arXiv",
    primaryClass = "hep-ph",
    reportNumber = "CERN-TH-2024-200, IPPP/24/73",
    month = "11",
    year = "2024"
}

@techreport{CMS_CDR,
      author        = "{CMS Collaboration}",
      title         = "{CMS Offline Software and Computing for HL-LHC Conceptual
                       Design Report}",
      institution   = "CERN",
      reportNumber  = "CERN-LHCC-2026-003, LHCC-G-186",
      address       = "Geneva",
      year          = "2026",
      url           = "https://cds.cern.ch/record/2957472",
}

@article{ATLAS_VplusJets,
	author = "{ATLAS Collaboration}",
	date = {2022/08/05},
	doi = {10.1007/JHEP08(2022)089},
	id = {Aad2022},
	isbn = {1029-8479},
	journal = {Journal of High Energy Physics},
	number = {8},
	pages = {89},
	title = {Modelling and computational improvements to the simulation of single vector-boson plus jet processes for the ATLAS experiment},
	url = {https://doi.org/10.1007/JHEP08(2022)089},
	volume = {2022},
	year = {2022}}
\end{document}